\documentclass[a4paper,onecolumn,unpublished]{quantumarticle}
\pdfoutput=1                      
\newif\ifSIappended
\SIappendedtrue
\ifSIappended\else
  \usepackage{xr-hyper}             
\fi
\usepackage[utf8]{inputenc}
\usepackage[english]{babel}
\usepackage[T1]{fontenc}
\usepackage{amsmath,amssymb}
\usepackage{graphicx}
\usepackage{booktabs,makecell}
\usepackage{tikz}
\usetikzlibrary{calc,decorations.pathreplacing,positioning,arrows.meta,shapes.geometric}
\usepackage{algorithm}
\usepackage{algpseudocode}

\usepackage[numbers,sort&compress]{natbib}
\usepackage{hyperref}
\AtBeginDocument{\normalfont}
\ifSIappended
  \newcommand{\siref}[1]{\ref{#1}}
\else
  \newcommand{\siref}[1]{\ref{si:#1}}
\fi
\graphicspath{{figures/}}
\begin{document}

\title{Parity Push Post-Selection with Soft Parity Constraints for Quantum LDPC Codes}

\author{Hong Jun Kim}
\affiliation{Department of Electrical and Computer Engineering, Korea University, Seoul 02841, Republic of Korea}
\author{Gun Sik Min}
\affiliation{Department of Electrical and Computer Engineering, Korea University, Seoul 02841, Republic of Korea}
\author{Seok-Hyung Lee}
\email{seokhyunglee@skku.edu}
\affiliation{Department of Quantum Information Engineering \& Department of Quantum Engineering, Sungkyunkwan University, Suwon 16419, Republic of Korea}
\author{Jun Heo}
\email{junheo@korea.ac.kr}
\affiliation{Department of Electrical and Computer Engineering, Korea University, Seoul 02841, Republic of Korea}

\maketitle

\begin{abstract}
Post-selection lowers the logical error rate of a quantum error-correcting code by rejecting unreliable shots.
For a code with $K$ logical observables, the logical gap, a standard post-selection criterion, involves $2^K$ logical classes, and existing approximations to it suffer from a non-convergence error floor or a shielding effect.
We introduce a decoder-agnostic post-selection strategy for quantum low-density parity-check (qLDPC) codes, called parity push post-selection (PPPS), which re-decodes a shot once per logical observable with a push toward the opposite parity.
Each push adds one check node and one variable node to the Tanner graph of the detector error model, and we accept a shot only when every re-decoding returns a solution in the original logical class.
We show that, with degenerate maximum-likelihood decoding, PPPS accepts a shot exactly when the logical gap exceeds the push strength $\lambda$.
Applied with the Relay-BP decoder to the $[[72,12,6]]$ bivariate bicycle code at $p=10^{-3}$ and the $[[144,12,12]]$ code at $p=3\times10^{-3}$ under circuit-level depolarizing noise, PPPS lowers the logical error rate by factors of about $50$ and $75$, respectively, at rejection rates near $1\%$.
We observe that PPPS rejects failures that prior post-selection strategies cannot reject, and thus shows improved performance.
\end{abstract}

\section{Introduction}

Post-selection rejects low-confidence shots and in return lowers the logical error rate (LER) of the accepted shots~\cite{aliferis2008accuracy,knill2005quantum,english2025thresholds}. 
For quantum low-density parity-check (qLDPC) codes~\cite{breuckmann2021quantum,bravyi2024high}, sacrificing approximately $1\%$ of shots yields a 10- to 100-fold reduction in LER at physical error rates around $10^{-3}$~\cite{lee2026efficient,wills2026forced,xie2026simple}.
Such a trade-off is beneficial wherever a rejected attempt can be discarded and repeated, most notably in the preparation of magic states~\cite{staples2026scalable,bombin2024fault,bravyi2005universal} and in lattice-surgery-based logical operations~\cite{akahoshi2026runtime,yoder2025tour}.
In fact, a circuit with $N_{\rm g}$ logical gates keeps a reasonable overall success probability only when the rejection rate per logical gate is at most of order $1/N_{\rm g}$~\cite{xie2026simple}, which makes LER suppression at low rejection rates particularly important.

The logical gap (or complementary gap), the log-likelihood difference between the two most probable logical classes consistent with the observed syndrome, is a standard criterion for post-selection~\cite{hutter2014efficient,gidney2025yoked,bombin2024fault,meister2024soft,smith2024mitigating}.
For a degenerate maximum-likelihood (DML) decoder, which returns the most probable logical class, the logical gap gives the failure probability of a shot through a logistic function (Eq.~\eqref{eq:pfail})~\cite{chen2025scalable}.
Thus, we can set a threshold on the gap and bound the failure probability of an accepted shot.
The surface code encodes one logical qubit, so one additional decoding run with flipped boundary conditions suffices to estimate the gap from the minimum-weight solutions of the two logical classes~\cite{gidney2025yoked,chen2025scalable}.
However, a qLDPC code with $K$ logical observables has $2^{K}$ logical classes, and computing the gap over all of them becomes infeasible as $K$ grows.

Three recent strategies avoid the $2^K$ enumeration.
Forced-gap post-selection (FG)~\cite{wills2026forced} re-decodes once per logical observable under a hard parity constraint with $K+1$ decoding runs in total, and estimates the logical gap from the resulting solutions.
Argument reweighting (AR)~\cite{xie2026simple} penalizes the initial solution, re-decodes once or twice, and accepts a shot when its logical class remains unchanged.
Cluster-based post-selection~\cite{lee2026efficient} quantifies decoding confidence from the cluster structure that clustering-based decoders produce, and rejects low-confidence shots.
It relies on clustering-based decoders such as the union-find decoder~\cite{delfosse2021almost} and belief propagation plus localized statistics decoding (BP+LSD)~\cite{hillmann2025localized}.
We instead build on FG and AR, which probe the logical gap by re-decoding with the same decoder, so that our strategy applies to a broad range of decoders.

Yet, both FG and AR have limitations acknowledged in the original papers~\cite{wills2026forced,xie2026simple}: a non-convergence error floor in the former and a shielding effect in the latter.
The floor in FG arises because the hard constraint prevents the forced runs from converging on most shots, including almost every correctly decoded one, so FG must accept such shots by convention, and failures among them pass any gap threshold.
AR suffers from the shielding effect because its soft penalty applies to a single solution, so other probable errors in the same logical class can shield errors of competing logical classes.

Table~\ref{tab:taxonomy} categorizes these schemes by what they manipulate (a physical correction or logical classes) and how they do it (hard constraint or soft penalty).
We propose parity push post-selection (PPPS), which applies a soft penalty to the logical sector of the initially decoded solution, that is, the $2^{K-1}$ logical classes that agree with it on a given logical observable.
For each logical observable, we perform a push by adding one auxiliary check and one slack variable to the detector error model (DEM)~\cite{gidney2021stim} Tanner graph. Here, we define the slack variable as an additional variable node that connects only to the auxiliary check.
Each auxiliary check requires the observable parity to flip, and its slack variable, with prior log-likelihood ratio (LLR) $\lambda$, allows this requirement to be violated at an LLR penalty of $\lambda$.
We call each re-decoding run under a push a probe, and we accept a shot only if all probes return to the logical class of the initial solution.
The push strength $\lambda$ is the only parameter of PPPS. 

\begin{table}[t]
\centering
\caption{\label{tab:taxonomy}Post-selection strategies for qLDPC codes, organized by what they manipulate (physical correction or logical classes) and how they do it (hard constraint or soft penalty).
Parity push post-selection occupies the soft-penalty/logical-classes cell.
As $\lambda\to\infty$ a parity push becomes the forced run of Ref.~\cite{wills2026forced}.}
\begin{tabular}{lcc}
\toprule
 & \textbf{Physical Correction} & \textbf{Logical Classes} \\
\midrule
\textbf{Hard Constraint} & --- & Forced gap \\
\textbf{Soft Penalty} & Argument reweighting & \textbf{Parity push} \\
\bottomrule
\end{tabular}
\end{table}

In principle, PPPS addresses both limitations of FG and AR.
Applying the penalty to an entire logical sector prevents another solution of the same logical class from shielding errors of a competing class.
Also, the initial solution still satisfies every check including the auxiliary check with only an LLR penalty of $\lambda$, so a probe can converge without finding a solution with the observable flipped.
Therefore, unlike in FG, we expect the probes of a correctly decoded shot to converge in most cases, so we can reject shots whose probes fail to converge.

Under DML decoding, all $K$ probes return to the initial logical class if and only if the logical gap over all $2^K$ logical classes exceeds $\lambda$.
PPPS therefore reduces a threshold on this gap to $K$ additional decoding runs and, through Eq.~\eqref{eq:pfail}, bounds the failure probability of every accepted shot.
DML decoding is itself intractable in general~\cite{iyer2015hardness}, so we run PPPS with heuristic BP-family decoders.
For such decoders $\lambda$ thresholds a weight difference rather than the exact gap, and we evaluate the resulting trade-off between the LER and the rejection rate numerically.

In this work, we formalize PPPS and evaluate it in simulations of the $[[72,12,6]]$ and $[[144,12,12]]$ bivariate bicycle codes~\cite{kovalev2013quantum,bravyi2024high} with the Relay-BP decoder~\cite{muller2025relay}.
The results show a $50$- to $75$-fold LER suppression at rejection rates of about $1\%$.
At the same decoding cost as FG, PPPS reaches LERs below the error floor of FG in every setting that we simulate and below or close to those of AR, which indicates that it addresses the limitations of both strategies.
Comparing two physical error rates on the $[[72,12,6]]$ code, we find that PPPS rejects failures increasingly through non-convergence as the error rate rises, which is the signal that FG cannot use.

Furthermore, we show that a decoder's ability to reject failures determines the LER, and its ability to return correctly decoded shots determines the rejection rate.
Numerical simulations on plain BP and belief propagation plus ordered statistics decoding (BP+OSD)~\cite{panteleev2021degenerate,roffe2020decoding} confirm this, demonstrating that PPPS is broadly compatible with various BP-family decoders and enhances their performance.

The main contributions of this work are:

(i) We introduce PPPS, which structurally addresses the limitations of previous logical-gap-based methods by adding one variable node and one check node to the DEM Tanner graph for each probe.

(ii) We provide numerical evidence of the effectiveness of PPPS on bivariate bicycle codes with the Relay-BP decoder, in particular large LER reductions at moderate rejection rates.

(iii) We show through numerical simulations that PPPS is applicable to BP-family decoders in general, and that the rejection rate depends on whether correctly decoded shots return under the push, and the LER on how many failures still return.

\section{Results}
\subsection{Detector error model and logical gap}\label{sec:dem}

The DEM~\cite{gidney2021stim} of a depolarizing circuit-level noise model consists of a detector matrix $H \in \mathbb{F}_2^{M \times N}$, an observable matrix $A \in \mathbb{F}_2^{K \times N}$, and a prior probability $p_i$ for each fault mechanism $i$.
Here $N$, $M$, and $K$ are the numbers of fault mechanisms, detectors, and logical observables, respectively.
A fault configuration $e = (e_1,\dots,e_N) \in \mathbb{F}_2^N$ produces the syndrome $\sigma = He$ and the logical outcome $c = Ae$.
The decoder has access only to $\sigma$ and returns a predicted error $\hat e$ satisfying $H\hat e = \sigma$, and decoding succeeds when $A\hat e = Ae$.

We assume independent fault mechanisms so that $P(e) = \prod_i p_i^{e_i}(1-p_i)^{1-e_i}$, where $p_i$ stands for the probability that fault mechanism $i$ occurs ($e_i=1$).
The weight
\[
  w(e) = \sum_{i \in \mathrm{supp}(e)} \ln\frac{1-p_i}{p_i} = -\ln P(e) + \sum_i \ln(1-p_i),
\]
where $\mathrm{supp}(e)=\{i: e_i=1\}$, differs from $-\ln P(e)$ by a term that sums over all fault mechanisms and does not depend on $e$.
Thus, the configuration of smallest weight is most probable.
Each term $\ln\frac{1-p_i}{p_i}$ is the prior LLR of the $i$-th fault mechanism, and a positive LLR favors $e_i = 0$, that is, the fault not occurring.

Two solutions $e$ and $e'$ of the same syndrome differ by $x=e+e' \in\ker H$, satisfying $Hx=0$, and they have different logical effects when $x$ changes the logical outcome, $Ax \neq 0$.
The solution set $\{e : He = \sigma\}$ therefore splits into $2^K$ logical classes according to the value of $c = Ae$.
We write $e^{(0)}$ for the solution that the initial decoding run, which we also call the baseline decoding, returns and $L^{(0)} = Ae^{(0)}$ for its logical class.
We define the mass of logical class $c$ as
\[
  Z_c(\sigma) = \sum_{e:\, He=\sigma,\, Ae=c} P(e),
\]
and the actual fault lies in logical class $c$ with probability $Z_c / \sum_{c'} Z_{c'}$, where the denominator is the total probability of observing $\sigma$.

Let $Z^{(1)} \geq Z^{(2)}$ be the two largest logical class masses, so that DML decoding returns the class of mass $Z^{(1)}$.
We define the logical gap as $g = \ln(Z^{(1)}/Z^{(2)})$, which coincides with the ``exact gap'' of Ref.~\cite{wills2026forced}.
In the literature, the logical (or complementary) gap more commonly denotes the weight difference between the minimum-weight solutions of the two classes~\cite{bombin2024fault,gidney2025yoked,smith2024mitigating}, which keeps only the largest term of each $Z_c$.
With our definition, DML decoding fails with probability
\begin{equation}
  \mathcal{P}(\mathrm{fail} \mid g, \alpha) = \frac{1+\alpha}{e^{g}+1+\alpha},
  \label{eq:pfail}
\end{equation}
where $\alpha$ is the total mass of the remaining $2^K-2$ classes divided by $Z^{(2)}$.
We derive Eq.~\eqref{eq:pfail} in Section~\ref{sec:gap}.
Setting $\alpha = 0$ recovers the two-class expression of Ref.~\cite{chen2025scalable}, derived for the free-energy difference between two homology classes of the toric code under DML decoding.
Equation~\eqref{eq:pfail} shows that the gap is more than a ranking score: if we accept shots satisfying $g > \lambda$, every accepted shot fails with probability below $(1+\alpha)/(e^{\lambda}+1+\alpha)$.
However, computing $g$ exactly requires the masses of all $2^K$ logical classes, which is infeasible for codes with large $K$.
We therefore aim to post-select on the logical gap without computing the mass of every logical class.

\subsection{Parity push for DML decoding}\label{sec:ml}    

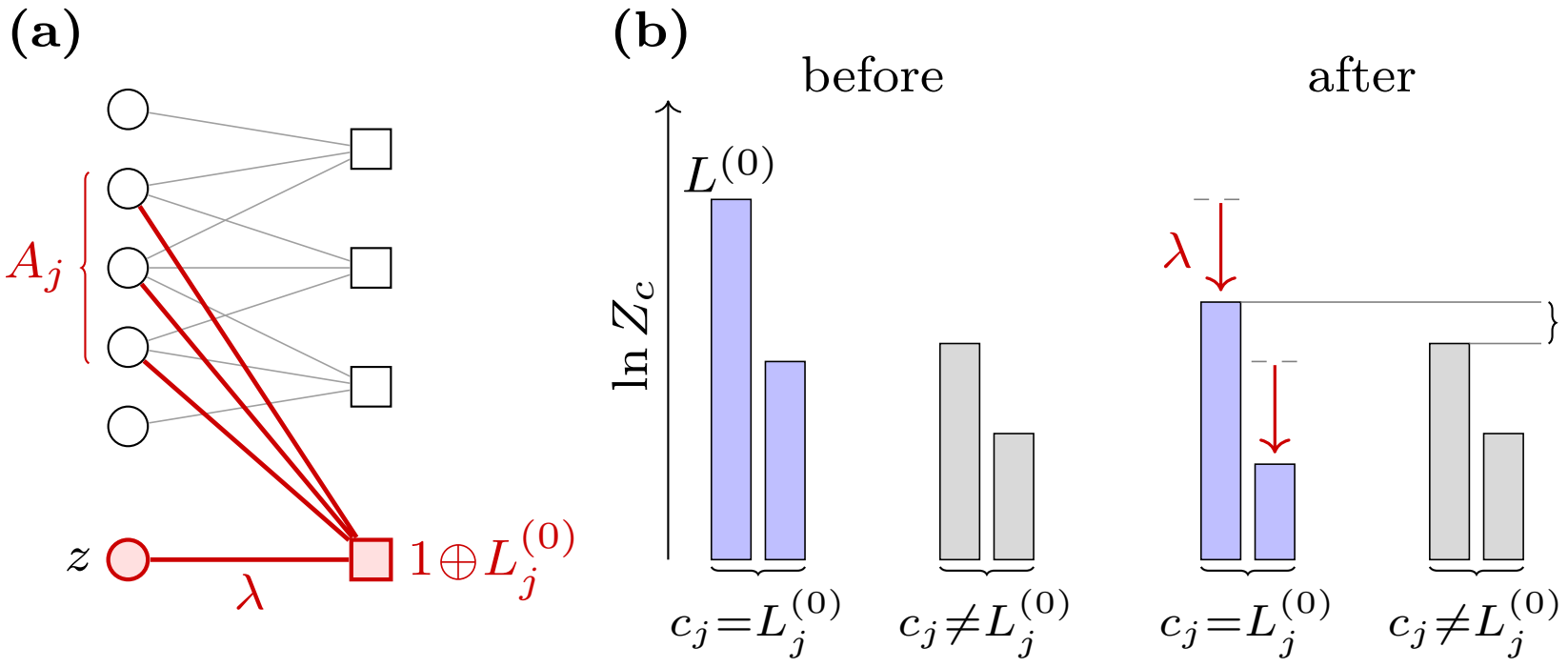
\begin{figure}[t]
\centering
\begin{tikzpicture}[
  font=\footnotesize,
  scale=1.25, every node/.style={transform shape},
  var/.style={circle,draw,minimum size=2.2mm,inner sep=0pt,fill=white},
  chk/.style={rectangle,draw,minimum size=2.2mm,inner sep=0pt,fill=white},
  aux/.style={draw=red!80!black,fill=red!12,thick},
  orig/.style={gray!75,line width=0.3pt},
  push/.style={red!80!black,line width=0.9pt},
  base/.style={fill=blue!25,draw,line width=0.3pt},
  opp/.style={fill=gray!30,draw,line width=0.3pt},
  sec/.style={anchor=north,inner sep=1pt,font=\scriptsize},
  ub/.style={decorate,decoration={brace,mirror,amplitude=1.5pt},line width=0.4pt}]

\begin{scope}
  \node[anchor=base west] at (-0.35,2.55) {\textbf{(a)}};
  \foreach \i/\y in {1/2.20,2/1.76,3/1.32,4/0.88,5/0.44}
    \node[var] (e\i) at (0.45,\y) {};
  \node[var,aux] (z) at (0.45,-0.30) {};
  \node[left=1pt of z,inner sep=1pt] {$z$};
  \foreach \i/\y in {1/1.98,2/1.32,3/0.66}
    \node[chk] (c\i) at (1.80,\y) {};
  \node[chk,aux] (h) at (1.80,-0.30) {};
  \draw[orig] (e1)--(c1) (e2)--(c1) (e3)--(c1);
  \draw[orig] (e2)--(c2) (e3)--(c2) (e4)--(c2);
  \draw[orig] (e3)--(c3) (e4)--(c3) (e5)--(c3);
  \draw[push] (e2)--(h) (e3)--(h) (e4)--(h);
  \draw[push] (z)--(h) node[midway,below=0.8pt,inner sep=1pt] {$\lambda$};
  \node[right=1pt of h,red!80!black,inner sep=1pt] {$1\!\oplus\!L^{(0)}_j$};
  \draw[decorate,decoration={brace,amplitude=1.5pt},red!80!black,line width=0.4pt]
    ($(e4.west)+(-0.10,-0.09)$) -- ($(e2.west)+(-0.10,0.09)$)
    node[midway,left=2pt,red!80!black,inner sep=1pt] {$A_j$};
\end{scope}

\begin{scope}[xshift=3.55cm]
  \node[anchor=base west] at (-0.55,2.55) {\textbf{(b)}};
  \draw[->,line width=0.4pt] (-0.10,-0.30) -- (-0.10,2.25);
  \node[rotate=90,anchor=south,inner sep=1pt] at (-0.14,0.95) {$\ln Z_c$};
  \node[anchor=base,inner sep=1pt] at (1.035,2.30) {before};
  \path[base] (0.14,-0.30) rectangle (0.36,1.70);
  \path[base] (0.44,-0.30) rectangle (0.66,0.80);
  \path[opp]  (1.41,-0.30) rectangle (1.63,0.90);
  \path[opp]  (1.71,-0.30) rectangle (1.93,0.40);
  \node[anchor=south,inner sep=1pt] at (0.25,1.70) {$L^{(0)}$};
  \draw[ub] (0.14,-0.34) -- (0.66,-0.34);
  \draw[ub] (1.41,-0.34) -- (1.93,-0.34);
  \node[sec] at (0.40,-0.42) {$c_j{=}L^{(0)}_j$};
  \node[sec] at (1.67,-0.42) {$c_j{\neq}L^{(0)}_j$};
  \node[anchor=base,inner sep=1pt] at (3.755,2.30) {after};
  \draw[dashed,gray,line width=0.3pt] (2.82,1.70) -- (3.10,1.70);
  \draw[dashed,gray,line width=0.3pt] (3.14,0.80) -- (3.40,0.80);
  \path[base] (2.86,-0.30) rectangle (3.08,1.13);
  \path[base] (3.16,-0.30) rectangle (3.38,0.23);
  \path[opp]  (4.13,-0.30) rectangle (4.35,0.90);
  \path[opp]  (4.43,-0.30) rectangle (4.65,0.40);
  \draw[->,red!80!black,line width=0.6pt] (2.97,1.68) -- (2.97,1.19);
  \draw[->,red!80!black,line width=0.6pt] (3.27,0.78) -- (3.27,0.29);
  \node[red!80!black,inner sep=0pt,anchor=east] at (2.82,1.42) {$\lambda$};
  \draw[ub] (2.86,-0.34) -- (3.38,-0.34);
  \draw[ub] (4.13,-0.34) -- (4.65,-0.34);
  \node[sec] at (3.12,-0.42) {$c_j{=}L^{(0)}_j$};
  \node[sec] at (4.39,-0.42) {$c_j{\neq}L^{(0)}_j$};
  \draw[gray,line width=0.3pt] (3.08,1.13) -- (4.75,1.13);
  \draw[gray,line width=0.3pt] (4.35,0.90) -- (4.75,0.90);
  \draw[decorate,decoration={brace,amplitude=1.5pt},line width=0.4pt]
    (4.79,1.13) -- (4.79,0.90);
\end{scope}
\end{tikzpicture}
\caption{The parity-push gadget.
(a) We add one auxiliary check (red square) and one slack variable $z$ (red circle, prior LLR $\lambda$) to the DEM Tanner graph.
The auxiliary check connects to the fault mechanisms in $\mathrm{supp}(A_j)$ (thick red edges) and carries the syndrome bit $1\oplus L^{(0)}_j$.
(b) Class masses before and after the push, drawn for four logical classes.
Blue bars show the classes of the baseline logical sector, which contains the baseline logical class $L^{(0)}$, and gray bars show the logical classes of the opposite logical sector.
Up to a factor common to all classes, the push lowers every class of the baseline logical sector by $\lambda$ and leaves the opposite sector unchanged [Eq.~\eqref{eq:mass}].
The probe returns if $L^{(0)}$ still exceeds the class of largest mass in the opposite sector [Eq.~\eqref{eq:return}].}
\label{fig:gadget}
\end{figure}

The key idea of PPPS is to penalize every solution that agrees with the initial solution on one logical observable and to check whether the decoder still returns the initial logical class. For each logical observable $j$, the value of $c_j$ splits the $2^K$ logical classes into two logical sectors of $2^{K-1}$ classes each. We call the sector $\{c : c_j = L^{(0)}_j\}$, which contains $L^{(0)}$, the baseline logical sector of observable $j$. We call the other sector, $\{c : c_j \neq L^{(0)}_j\}$, the opposite logical sector.
We add one auxiliary check and one slack variable $z$ to the DEM Tanner graph for each logical observable $j$ to push the entire baseline logical sector [Fig.~\ref{fig:gadget}(a)]. 
The detector matrix $H$ transforms to 
\begin{equation}
  H' = \begin{bmatrix} H & 0 \\ A_j & 1 \end{bmatrix},
  \label{eq:Hprime}
\end{equation}
where the new row consists of the $j$-th row $A_j$ of $A$ and an additional $1$.
The row $A_j$ marks the fault mechanisms that flip the $j$-th observable, and the additional $1$ connects the slack variable to the auxiliary check.
We append the bit $1 \oplus L^{(0)}_j$ to the syndrome as the value of the auxiliary check, so that the check enforces $A_j e \oplus z = 1 \oplus L^{(0)}_j$.
The slack variable $z$ connects only to the auxiliary check and carries the prior probability $p_z = 1/(1+e^{\lambda})$, corresponding to a prior LLR of $\lambda \geq 0$. In DEM terms, $z$ is one more fault mechanism with probability $p_z$, and it corresponds to no physical fault.
Other edge connections and prior LLR values remain unchanged.

Since the auxiliary check enforces $A_j e \oplus z = 1 \oplus L^{(0)}_j$, the value of the slack variable $z$ is determined by the resulting error configuration $e$: a configuration that retains the baseline parity ($A_je=L_j^{(0)}$) corresponds to $z=1$, while a configuration that flips it ($A_je\neq L_j^{(0)}$) corresponds to $z=0$.
As $P(z{=}1) = e^{-\lambda} P(z{=}0)$ by the definition of the LLR, every $e$ with $He = \sigma$ has probability $P'(e)=P(e)P(z)$ in the deformed Tanner graph:
\begin{equation}
  P'(e) =
  \begin{cases}
    \dfrac{1}{1+e^{\lambda}} P(e) & \text{if } A_j e = L^{(0)}_j, \\[8pt]
    \dfrac{e^{\lambda}}{1+e^{\lambda}} P(e) & \text{if } A_j e \neq L^{(0)}_j.
  \end{cases}
  \label{eq:reweight}
\end{equation}
Equation~\eqref{eq:reweight} holds exactly, and we derive Eqs.~\eqref{eq:reweight}--\eqref{eq:threshold} in Section~\ref{sec:gap}.

Thus, the mass of logical class $c$ becomes
\begin{equation}
  Z'_c =
  \begin{cases}
    \dfrac{1}{1+e^{\lambda}} Z_c & \text{if } c_j = L^{(0)}_j, \\[8pt]
    \dfrac{e^{\lambda}}{1+e^{\lambda}} Z_c & \text{if } c_j \neq L^{(0)}_j.
  \end{cases}
  \label{eq:mass}
\end{equation}
Equation~\eqref{eq:mass} shows that the mass of every logical class in the baseline logical sector of observable $j$ decreases by the factor $e^{-\lambda}$ relative to the classes of the opposite logical sector, that is, by exactly $\lambda$ in log-mass.
The remaining factor $e^{\lambda}/(1+e^{\lambda})$ is common to all classes and does not change their order.
Thus, unlike AR, which pushes a single solution, our gadget pushes an entire logical sector by $\lambda$ nats relative to the opposite logical sector and thereby avoids the shielding effect.
Figure~\ref{fig:gadget}(b) illustrates the class masses before and after the push. 

We regard a shot as reliable when the probes for all logical observables return to the logical class $L^{(0)}$ under the push.
If a DML decoder performs both the initial decoding and the probes, Eq.~\eqref{eq:mass} directly gives the outcome of each probe.
Under the assumption that the class of largest mass is unique in each decoding run, the initial decoding step returns logical class $L^{(0)}$, and the probe on logical observable $j$ returns to that class if and only if
\begin{equation}
  e^{-\lambda} Z_{L^{(0)}} > \max_{c:\, c_j \neq L^{(0)}_j} Z_c .
  \label{eq:return}
\end{equation}
Every logical class other than $L^{(0)}$ differs from it in at least one logical observable, so the $K$ opposite logical sectors cover all $2^K-1$ competing classes.
Hence,
\begin{equation}
  \text{all probes return}
  \iff e^{-\lambda} Z^{(1)} > Z^{(2)}
  \iff g > \lambda .
  \label{eq:threshold}
\end{equation}
Therefore, under DML decoding, $K$ additional decoding steps place a threshold on the gap over $2^K$ logical classes, and the push strength $\lambda$ is that threshold.

We note two limiting cases of the construction: $\lambda=0$ and $\lambda=\infty$.
At $\lambda=0$ Eq.~\eqref{eq:reweight} gives $P'(e) \propto P(e)$, so the deformed problem remains equivalent to the initial problem.
For $\lambda=\infty$, the slack variable $z$ can no longer take the value $1$.
Consequently, the auxiliary check becomes a hard parity constraint, and the probes become forced runs of Ref.~\cite{wills2026forced}.

\subsection{Parity push for practical decoders}\label{sec:practical} 

DML decoding remains impractical for qLDPC codes, so we first adapt our strategy to a minimum-weight decoder, which ignores degeneracy.
A minimum-weight decoder returns the minimum-weight solution $e^{(0)}$~\cite{fuentes2021degeneracy}, and we take its logical class $L^{(0)}$ as the most probable one.
After the initial decoding returns $e^{(0)}$, the $j$-th probe increases the weight of every solution in the baseline logical sector of observable $j$ by $\lambda$.
A probe therefore flips only if the minimum-weight solution of the opposite logical sector has weight below $w(e^{(0)}) + \lambda$, and all probes return if and only if the minimum weight outside class $L^{(0)}$ exceeds $w(e^{(0)})$ by more than $\lambda$.
In this case, $\lambda$ thresholds the weight difference between the minimum-weight solutions inside and outside class $L^{(0)}$ rather than the exact logical gap, so PPPS becomes equivalent to post-selection on the conventional logical gap~\cite{bombin2024fault,gidney2025yoked,smith2024mitigating}.
Thus, Eq.~\eqref{eq:pfail} no longer converts the threshold into a strict bound on the failure probability and gives only an approximation.

Exact minimum-weight decoding is also intractable for general qLDPC codes, so we use practical BP-family decoders.
Such a decoder is heuristic: within its iteration budget it returns a solution that typically has low weight, with no guarantee that the weight is minimal.
Practical decoders therefore give no guarantee on the failure probability, and the threshold becomes an even rougher approximation than with a minimum-weight decoder. 

\begin{figure}[t]
\centering
\begin{tikzpicture}[
  font=\footnotesize, >=Stealth,
  blk/.style={draw,line width=0.5pt,fill=white,align=center,inner sep=4pt,minimum height=7.5mm},
  gad/.style={blk,draw=red!80!black,line width=0.7pt,align=left,inner sep=5pt},
  card/.style={draw,line width=0.5pt,fill=white,minimum width=15mm,minimum height=7.5mm,inner sep=0pt},
  dec/.style={draw,line width=0.5pt,fill=white,diamond,aspect=2,align=center,inner sep=0.5pt},
  rej/.style={blk,fill=black!10,font=\footnotesize\bfseries,minimum width=14mm},
  acc/.style={blk,line width=1.1pt,font=\footnotesize\bfseries,minimum width=14mm},
  lab/.style={font=\scriptsize,inner sep=1.5pt},
  arr/.style={->,line width=0.6pt},
  ph/.style={font=\scriptsize\scshape,text=black!60,anchor=south}]
\node (sig) {$\sigma$};
\node[blk, right=5mm of sig] (base) {baseline\\decoder};
\node[gad, right=15mm of base] (gad)
 {$H'_j \leftarrow \begin{bmatrix} H & 0\\ A_j & 1\end{bmatrix}$\\[3pt]
  $\sigma'_j \leftarrow \begin{bmatrix} \sigma\\ 1\oplus L^{(0)}_j\end{bmatrix}$\\[3pt]
  $p_z \leftarrow 1/(1+e^{\lambda})$};
\coordinate (cw) at ([xshift=8mm]gad.east);
\node[card] (c3) at ($(cw)+(0.75+0.26,0.26)$) {};
\node[card] (c2) at ($(cw)+(0.75+0.13,0.13)$) {};
\node[card] (c1) at ($(cw)+(0.75,0)$) {probe $j$};
\node[lab,below=1pt of c1] (par) {$j=1,\dots,K$ in parallel};
\node[dec, right=9mm of c1] (d2) {all $K$\\return?};
\node[acc, right=5mm of d2] (acc) {Accept};
\node[rej, below=8mm of base] (rej1) {Reject};
\node[rej, below=8mm of d2] (rej2) {Reject};
\draw[arr] (sig) -- (base);
\draw[arr] (base) -- node[lab,above] {$e^{(0)},\,L^{(0)}$} (gad);
\draw[arr] (base) -- node[lab,right,align=left] {not\\converged} (rej1);
\draw[arr] (gad) -- (c1);
\draw[arr] (c1) -- (d2);
\draw[arr] (d2) -- node[lab,above] {yes} (acc);
\draw[arr] (d2) -- node[lab,right] {no} (rej2);
\node[blk,inner sep=4pt,anchor=north west] (leg) at ($(gad.south west)+(0,-0.45)$)
 {\renewcommand{\arraystretch}{1.15}%
  \begin{tabular}{@{}l@{\hspace{6pt}}l@{}}
   \multicolumn{2}{@{}l@{}}{outcome of probe $j$:}\\
   \textbf{return}  & converged, $A\hat e^{(j)} = L^{(0)}$\\
   \textbf{flip}    & converged, $A\hat e^{(j)} \neq L^{(0)}$\\
   \textbf{abstain} & not converged
  \end{tabular}};
\path (gad.north) ++(0,0.3) coordinate (by);
\draw[decorate,decoration={brace,amplitude=3.5pt},line width=0.5pt,black!60]
  (sig.west |- by) -- (base.east |- by) node[ph,midway,yshift=4pt] {Phase 1 --- baseline run};
\draw[decorate,decoration={brace,amplitude=3.5pt},line width=0.5pt,black!60]
  (gad.west |- by) -- (c3.east |- by) node[ph,midway,yshift=4pt] {Phase 2 --- $K$ parity-push probes};
\draw[decorate,decoration={brace,amplitude=3.5pt},line width=0.5pt,black!60]
  (d2.west |- by) -- (acc.east |- by) node[ph,midway,yshift=4pt] {Phase 3 --- decision};
\end{tikzpicture}
\caption{Overview of parity push post-selection.
Phase 1: the decoder runs on the syndrome $\sigma$ and returns a baseline correction $e^{(0)}$ with logical class $L^{(0)}=Ae^{(0)}$.
The shot is rejected at once if the baseline decoding does not converge.
Phase 2: for each of the $K$ logical observables, a probe decodes the modified problem $(H'_j,\sigma'_j)$ with one auxiliary check and one slack variable of prior LLR $\lambda$ [Fig.~\ref{fig:gadget}(a)].
The probes need only $L^{(0)}$ and therefore run independently of each other.
Phase 3: each probe returns, flips or abstains, and the shot is accepted only if all $K$ probes return.
As $\lambda\to\infty$ each probe becomes a forced run of Ref.~\cite{wills2026forced}.}
\label{fig:overview}
\end{figure}
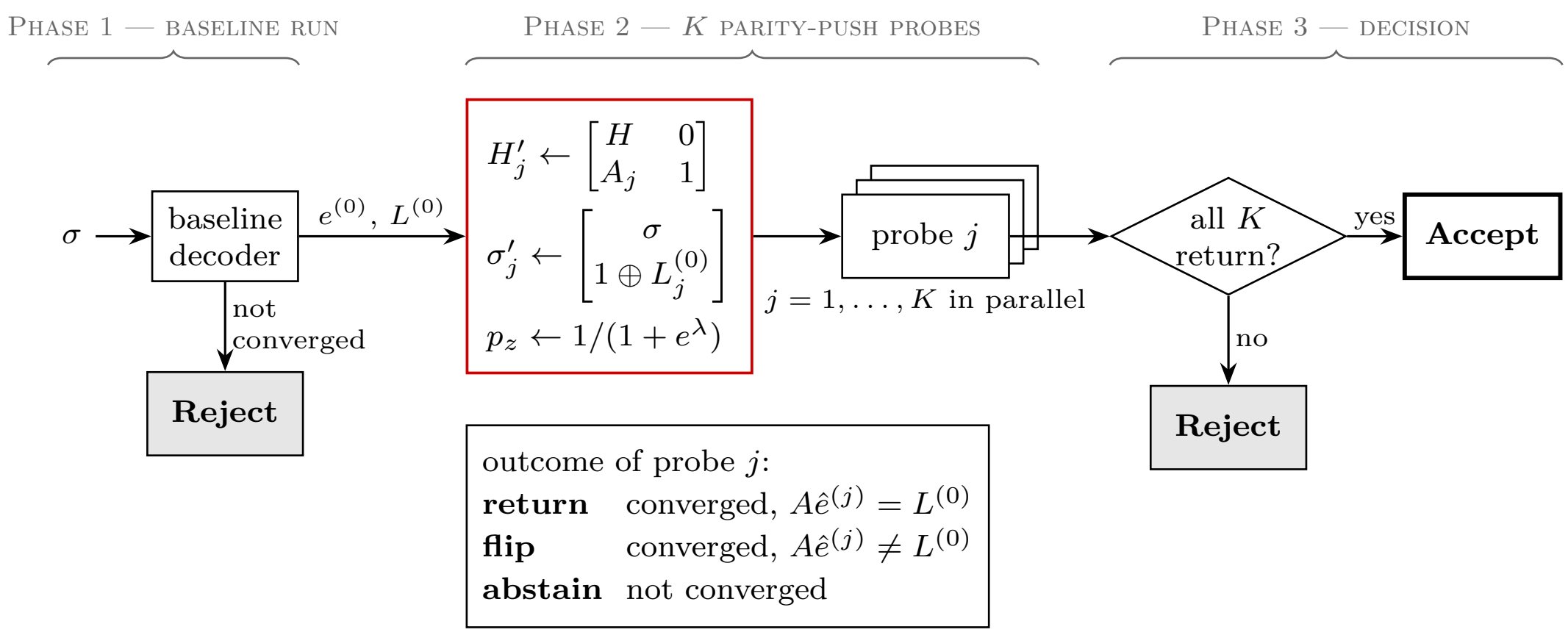

The slack variable and the auxiliary check are ordinary nodes of the Tanner graph, so any decoder that returns a correction satisfying the syndrome on this Tanner graph can run the probes.
Within its budget, such a decoder may fail to find the minimum-weight solution of a logical sector, or even to find any solution at all~\cite{poulin2008iterative,raveendran2021trapping}.
A practical probe therefore has three possible outcomes: it may \emph{return} if it converges to a solution in class $L^{(0)}$, \emph{flip} if it converges to a solution in a class other than $L^{(0)}$, or \emph{abstain} if it does not converge within its budget.
We accept a shot only if every probe \emph{returns}, and reject it if any probe \emph{flips} or \emph{abstains}.
While we do not read the hard decision of an abstaining probe, Section~\ref{sec:relaxed} evaluates a relaxed rule that does.
Figure~\ref{fig:overview} summarizes the overall procedure of our strategy, and Algorithm~\ref{alg:ppps} in Section~\ref{sec:alg} gives its pseudocode.

We add three remarks on the rule.
First, we compare the full logical class $A\hat e^{(j)}$ of each probe solution with $L^{(0)}$ and not only bit $j$.
A probe can converge to a class that keeps bit $j$ of $L^{(0)}$ but differs in another observable.
Such a probe does not support $L^{(0)}$, so we count it as a flip.
Second, we push every observable with the same $\lambda$.
Each probe depends on $(\sigma, L^{(0)}, j, \lambda)$, and the probes are independent and differ only in the observable $j$.
Third, we run each probe from a cold start: we initialize all messages from the priors and carry over no messages from the initial run, sharing only its seed (Section~\ref{sec:impl}).
Since each probe is independent, the decision does not depend on the order in which the probes run.

\emph{Relation to the forced run.}
When the baseline run and the forced runs find only one logical class, which requires that no forced run converges, FG sets the gap to infinity by convention and accepts the shot~\cite{wills2026forced}.
A forced run under a hard constraint fails to converge on almost every correctly decoded shot (Section~\ref{sec:baselines}), so rejecting non-converged shots instead would reject nearly all correctly decoded shots.
Failures with an infinite gap therefore pass for any choice of the gap threshold, which produces the error floor.
In PPPS the initial solution with $z=1$ still satisfies every check including the auxiliary check, and its weight increases by only $\lambda$.
With this solution available, a probe tends to converge on correctly decoded shots, and we can treat one that fails to converge as unreliable and reject the shot.
Consequently, PPPS can reject failures that FG accepts.
This signal has a price: a correctly decoded shot whose probe fails to converge under the push is also rejected, and we expect this cost to grow with $\lambda$.
Whether the advantage or the cost dominates depends on how often correctly decoded shots and failures return under the push, which we examine numerically via simulation.

\emph{Computational cost.}
PPPS runs the decoder $K+1$ times per shot.
The probes start after the initial decoding, and they are independent of each other.
If they run in parallel, the latency is that of two decoding stages, as for FG~\cite{wills2026forced}.
Each probe adds one row and one column to the check matrix, which raises the number of edges of the Tanner graph by about $2\%$.
A sequential implementation can stop at the first flip or abstention, but this saves runs only on rejected shots.
An accepted shot always runs all $K$ probes.
Section~\ref{sec:setup} gives the resulting cost for our decoder budgets, and Section~\ref{sec:cost} derives it.

\subsection{Simulation setup}\label{sec:setup}
We benchmark our post-selection strategy via simulations on bivariate bicycle codes~\cite{bravyi2024high} with BP-family decoders.
The simulation consists of Z-basis memory experiments whose $K=12$ logical observables are the logical Z operators.
We simulate the $[[72,12,6]]$ and $[[144,12,12]]$ codes with 6 and 12 rounds of syndrome extraction, respectively, under circuit-level depolarizing noise of strength $p$ (Section~\ref{sec:setup-methods}).
We consider three settings: the $[[72,12,6]]$ code at $p=10^{-3}$ and $3\times10^{-3}$ and the $[[144,12,12]]$ code at $p=3\times10^{-3}$ (Fig.~\ref{fig:relay}). We report the LER per round and the rejection rate per shot, as in Ref.~\cite{wills2026forced}.
A shot of the $[[144,12,12]]$ code runs twice as many rounds as a shot of the $[[72,12,6]]$ code, so the rejection rates per shot of the two codes are not directly comparable.

We use Relay-BP~\cite{muller2025relay} to compare PPPS with FG and AR.
All three methods run on the same shots, start from the same initial solution, and use the same decoder settings for every re-decoding run.
The only exception is the $[[144,12,12]]$ code, where PPPS runs on fewer correctly decoded shots than FG and AR (Section~\ref{sec:stats}).
For PPPS, the probes use a smaller Relay-BP budget than the baseline, with the same parameters for every probe so that all probes apply the same push.
Table~\ref{tab:decoders} lists the parameter values, and Section~\ref{sec:impl} describes the implementation of the probes.
With these budgets, PPPS runs about $2$ and $5$ times the operations of the baseline decoding on the $[[72,12,6]]$ and $[[144,12,12]]$ codes, respectively.
Its latency with parallel probes is about $1.1$ and $1.3$ times that of the baseline (Section~\ref{sec:cost}).

Running each post-selection method on every shot is infeasible, so we evaluate the performance on two sets of shots.
The first set holds the failures of the baseline decoder, which we collect with a two-pass procedure for the $[[72,12,6]]$ code at $p=10^{-3}$ and the $[[144,12,12]]$ code and with the full baseline decoder for the $[[72,12,6]]$ code at $p=3\times10^{-3}$ (Section~\ref{sec:stats}).
The two-pass procedure gives lower bounds on the numbers of failures and the baseline LERs, and we run each method on every collected failure.
The second set is a sample of correctly decoded shots, on which we run each method and estimate the fraction that it rejects.
The error bars on the rejection rate therefore reflect the sample of correctly decoded shots, and those on the LER reflect the number of accepted failures (Section~\ref{sec:stats}).

\subsection{Results for Relay-BP on the [[72,12,6]] code}\label{sec:relay}       

\begin{figure*}[t]
\centering
\includegraphics[width=\textwidth]{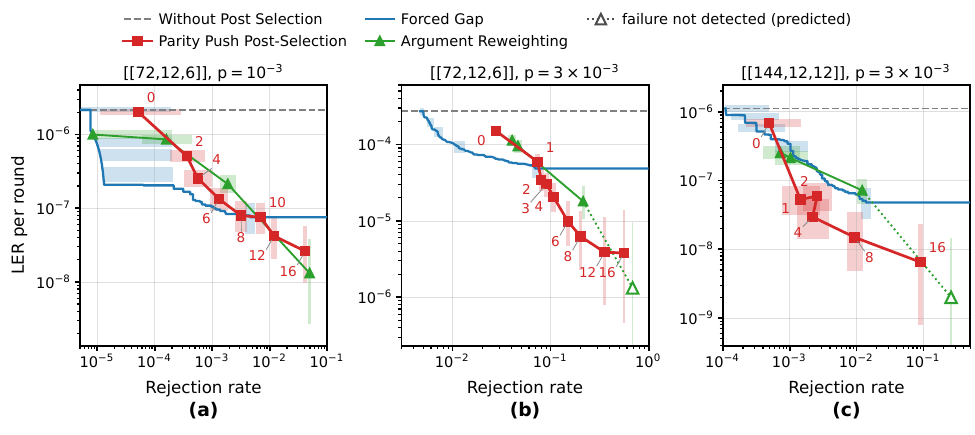}
\caption{Post-selection with Relay-BP on bivariate bicycle codes.
Logical error rate per round against rejection rate per shot for the $[[72,12,6]]$ code at (a) $p=10^{-3}$ and (b) $p=3\times10^{-3}$ with six rounds, and (c) the $[[144,12,12]]$ code at $p=3\times10^{-3}$ with twelve rounds.
The dashed line is the logical error rate without post-selection ($2.17\times10^{-6}$, $2.71\times10^{-4}$ and $1.12\times10^{-6}$).
PPPS points are labeled by the push strength $\lambda$.
The FG curve sweeps the gap threshold and stays flat beyond the point where every finite gap is rejected.
AR points sweep the reweighting strength $z$. 
Open markers show points with no accepted failure.
They are placed at the logical error rate that half a failure would give, and their bars extend to the upper confidence limit.
The rejection rate includes the shots on which the baseline decoder does not converge, which is the same for all three methods.
Shaded regions are 95\% confidence intervals (Wilson for the rejection rate, Garwood for the logical error rate~\cite{wilson1927probable,garwood1936fiducial}).
Correctly decoded shots are sampled: $19{,}999$ for all methods in (a), $9{,}936$ for all methods in (b), and $5{,}292$ for PPPS and $9{,}890$ for FG and AR in (c).}
\label{fig:relay}
\end{figure*}

Figure~\ref{fig:relay}(a) and (b) present the simulation results for the $[[72,12,6]]$ code at $p = 10^{-3}$ and $p=3\times10^{-3}$, respectively.
Each panel plots the LER per round against the rejection rate for FG, AR, and PPPS, all with Relay-BP.
We obtain the FG and AR curves from our own implementations with the decoder settings given in Methods, so their values differ from those reported in Refs.~\cite{wills2026forced,xie2026simple}.
Supplementary Tables~\siref{stab:ppps}--\siref{stab:ar} list the values of every point.

In (a) the PPPS curve starts close to the baseline at $\lambda=0$, which serves as a control without push (Section~\ref{sec:impl}).
The LER then decreases with $\lambda$ to about $50$ times below the baseline at a rejection rate of about $1\%$.
At the lowest rejection rates FG lies below PPPS, and the two curves meet near a rejection rate of $0.3\%$.
The FG curve then flattens at a rejection rate of about $0.5\%$, where every failure with a finite gap is rejected.
PPPS continues below this floor and reaches a two to three times lower LER at rejection rates above $1\%$.
AR stays within a factor of about two of PPPS over the whole range.

In (b) the PPPS curve starts at a rejection rate of $2.8\%$, since the probes without a push already reject about $2\%$ of correctly decoded shots (Section~\ref{sec:impl}).
The LER then decreases with $\lambda$ and reaches a value $43$ times below the baseline at a rejection rate of $20\%$.
FG lies below PPPS until its curve flattens at a rejection rate of $7\%$, and PPPS falls below this floor from a rejection rate of $8\%$.
AR needs about twice the rejection rate of PPPS for a similar LER.

PPPS rejects more failures as $\lambda$ increases at both physical error rates, but the two rates differ in how these failures are rejected, by a flip or by an abstention.
For example, at $\lambda=8$ only $3.6\%$ of failures are rejected through abstention at $p=10^{-3}$, while this proportion grows to $21\%$ at $p=3\times10^{-3}$.
We attribute this rise in abstention to the decoder needing more iterations to converge at the higher physical error rate (Supplementary Note~\siref{snote:origin}).
PPPS rejects a shot whose probes do not converge, whereas FG accepts most of these failures because their forced runs do not converge either (Section~\ref{sec:practical}).
Since abstention rejects a larger share of the failures at $p=3\times10^{-3}$, PPPS falls below the FG floor with a smaller push at the higher physical error rate.

\subsection{Results for Relay-BP on the [[144,12,12]] code}\label{sec:relay144}

In Fig.~\ref{fig:relay}(c) the PPPS curve starts at a rejection rate of $0.05\%$, and the probes without a push reject only $2$ of the $5{,}292$ sampled correctly decoded shots.
A push of $\lambda=1$ already lowers the LER by about $20$ times at a rejection rate of $0.14\%$.
The LER then decreases further with $\lambda$, and the slight increase at $\lambda=2$ lies within the statistical uncertainty.
Simulation results show that a rejection rate of $1\%$ lowers the LER by about $75\times$. At $\lambda=16$ the reduction reaches about $170\times$ at a rejection rate of $9\%$, but only two failures remain accepted (Supplementary Table~\siref{stab:ppps}).

Unlike for the $[[72,12,6]]$ code, PPPS lies below FG at every rejection rate from about $0.1\%$.
The FG curve flattens at a rejection rate of $1.4\%$, and PPPS falls below this floor already at $0.2\%$.
AR stays above PPPS at every rejection rate from about $0.1\%$, by a factor of about five near $1\%$.

At $p=3\times10^{-3}$, the $[[144,12,12]]$ code differs from the $[[72,12,6]]$ code in how PPPS rejects the failures.
On the $[[144,12,12]]$ code, PPPS rejects most failures by a flip, and abstention accounts for $5\%$ of the failures at $\lambda=8$ against $21\%$ on the $[[72,12,6]]$ code.
The probes also rarely abstain on correctly decoded shots of the $[[144,12,12]]$ code, which keeps the rejection rate low.
Part of this difference may come from the probe budget.
Because the DEM Tanner graph of the $[[144,12,12]]$ code is larger, we set its probe budget about four times as large as that of the $[[72,12,6]]$ code (Table~\ref{tab:decoders}).
These data therefore do not separate the effect of the code from that of the budget (Supplementary Note~\siref{snote:origin}).

\subsection{Results for other BP-family decoders}\label{sec:decoders} 

\begin{figure}[t]
\centering
\includegraphics[width=1\textwidth]{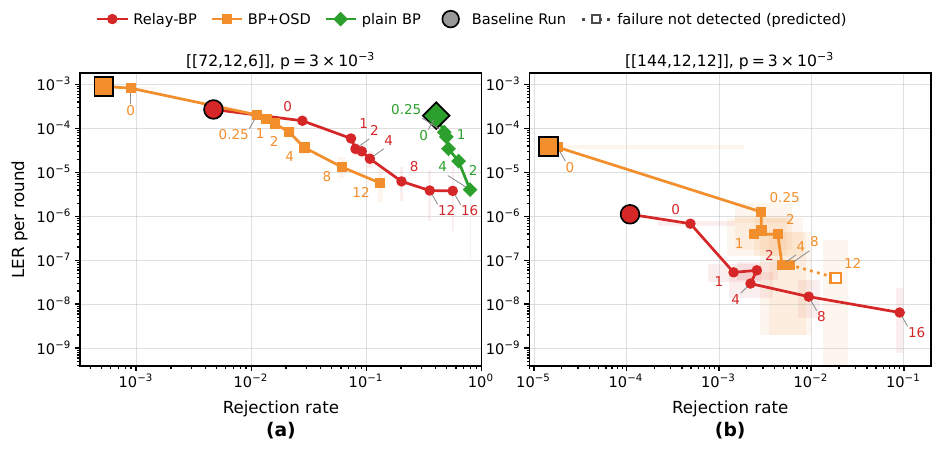}
\caption{PPPS with three BP-family decoders at $p=3\times10^{-3}$.
Logical error rate per round against rejection rate per shot for (a) the $[[72,12,6]]$ code, where all decoders run on the same 200{,}000 shots, and (b) the $[[144,12,12]]$ code, where BP+OSD runs on all 1.06 million shots and Relay-BP on 28.2 million.
Filled markers are the baseline points of each decoder: the rejection rate is the fraction of shots on which the baseline decoder does not converge, and the logical error rate is that of the remaining shots.
PPPS points are labeled by $\lambda$.
Plain BP is not shown in (b) because it converges on only 30.3\% of shots and no failures occur among them (Table~\ref{tab:samples}).
Open markers show points with no accepted failure.
They are placed at the logical error rate that half a failure would give, and their bars extend to the upper confidence limit.
Shaded regions are 95\% confidence intervals (Wilson for the rejection rate, Garwood for the logical error rate).
The Relay-BP probes use a budget of 25 legs in (a) and 100 legs in (b) against 300 for the baseline (Table~\ref{tab:decoders}).}
\label{fig:decoders}
\end{figure}

Figure~\ref{fig:decoders} presents the results of PPPS with three BP-family decoders: plain BP, BP+OSD, and Relay-BP, on the $[[72,12,6]]$ code (a) and the $[[144,12,12]]$ code (b), both at $p = 3\times10^{-3}$.
Plain BP runs min-sum decoding for 100 iterations, BP+OSD~\cite{panteleev2021degenerate,roffe2020decoding} adds a combination-sweep OSD post-processing of order 10, and Relay-BP uses the settings of Table~\ref{tab:decoders}.
Each curve starts from the baseline point of its own decoder, which rejects only the shots on which the decoder does not converge.
In (a) the baseline of BP+OSD rejects no shot, that of Relay-BP rejects $0.5\%$, and that of plain BP already rejects $40\%$.
The low starting LER of plain BP therefore reflects the shots it excludes rather than better decoding.
Strengthening the push $\lambda$ lowers the LER and in return raises the rejection rate for every decoder.
This trend comes from two factors: fewer failures return, which lowers the LER, and more correctly decoded shots fail to return, which raises the rejection rate (Section~\ref{sec:stats}).
Supplementary Table~\siref{stab:decoders} lists the values of every point.

In (a), we observe that BP+OSD outperforms Relay-BP in that it requires a lower rejection rate to achieve an equivalent LER.
While Relay-BP requires an $8.0\%$ rejection rate to reach an LER of $3.5\times10^{-5}$, BP+OSD requires only $2.9\%$ to reach a similar level.
Plain BP reaches this LER with a rejection rate of $52\%$, which is far higher.
This comparison holds for our decoder settings, in which the Relay-BP probes use a smaller budget than their baseline whereas the probes of BP+OSD and plain BP use the settings of their baselines (Section~\ref{sec:impl}).
For all three decoders, the LER decreases consistently as $\lambda$ increases.

The $[[144,12,12]]$ simulation results in (b) also show that both BP+OSD and Relay-BP lower the LER with PPPS.
BP+OSD reaches an LER of $7.9\times10^{-8}$ at a rejection rate of $0.48\%$, while Relay-BP reaches $1.5\times10^{-8}$ at a rejection rate of $0.94\%$.
At $\lambda=12$ BP+OSD accepts none of the $497$ failures, so resolving the LER at larger $\lambda$ would require more shots.
We also note that we do not post-select plain BP on the $[[144,12,12]]$ code because it converges on only $30.3\%$ of the shots and no failure occurs among them (Table~\ref{tab:samples}).

\subsection{Results for the relaxed rejection rule}\label{sec:relaxed} 

\begin{figure*}[t]
\centering
\includegraphics[width=\textwidth]{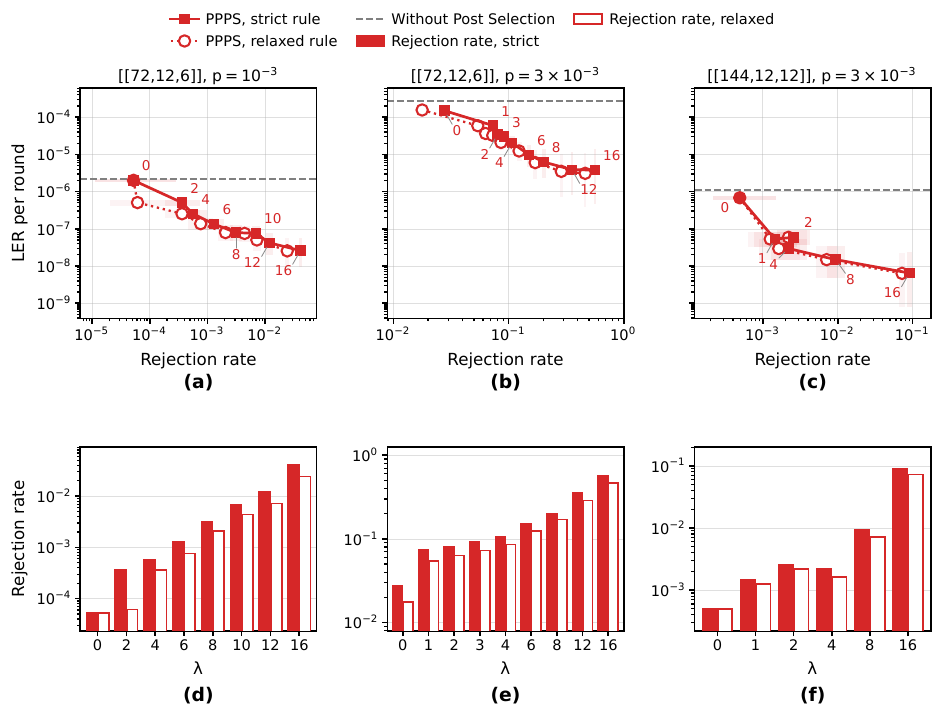}
\caption{Strict and relaxed rejection rules of PPPS with Relay-BP.
Logical error rate per round against rejection rate per shot for the $[[72,12,6]]$ code at (a) $p=10^{-3}$ and (b) $p=3\times10^{-3}$ with six rounds, and (c) the $[[144,12,12]]$ code at $p=3\times10^{-3}$ with twelve rounds.
Panels (d)--(f) show the rejection rate per shot of both rules at each push strength $\lambda$ for the settings of (a)--(c), respectively.
The strict rule (solid lines, filled bars) rejects a shot when any probe flips or abstains, as in Fig.~\ref{fig:relay}.
The relaxed rule (dotted lines, open bars) reads the hard decision of each non-converged probe and accepts the shot when no probe flips and every such decision lies in $L^{(0)}$.
In (a)--(c) strict points are labeled by the push strength $\lambda$, and each relaxed point has the same $\lambda$ as the strict point it pairs with.
The dashed line is the logical error rate without post-selection ($2.17\times10^{-6}$, $2.71\times10^{-4}$ and $1.12\times10^{-6}$).
The rejection rate includes the shots on which the baseline decoder does not converge.
Shaded regions in (a)--(c) are 95\% confidence intervals (Wilson for the rejection rate, Garwood for the logical error rate).
The bars in (d)--(f) are point estimates (Supplementary Tables~\siref{stab:ppps} and~\siref{stab:relaxed}).
Both rules run on the same failures and the same sample of correctly decoded shots: $19{,}999$ in (a) and (d), $9{,}936$ in (b) and (e), and $5{,}292$ in (c) and (f).}
\label{fig:relaxed}
\end{figure*}

Figure~\ref{fig:relaxed} compares the strict rejection rule of PPPS with a relaxed one for the three settings of Fig.~\ref{fig:relay}.
The strict rule rejects a shot whose probe abstains, whereas the relaxed rule reads the hard decisions of the non-converged probes and accepts the shot when no probe flips and every decision lies in $L^{(0)}$.
For $\lambda \geq 2$ the relaxed curve lies to the left of the strict curve in panels (a)--(c).
The relaxed rule lowers the rejection rate while the LER stays within the confidence interval of the strict rule: at $\lambda = 16$ it drops from $4.1\%$ to $2.4\%$ for the $[[72,12,6]]$ code at $p = 10^{-3}$, from $56\%$ to $46\%$ for the same code at $p = 3\times10^{-3}$, and from $9.0\%$ to $7.2\%$ for the $[[144,12,12]]$ code.
Supplementary Table~\siref{stab:relaxed} lists the values of every point.

The gain comes from correctly decoded shots, while failures are almost untouched: the relaxed rule accepts at most two additional failures at any $\lambda$ for the $[[72,12,6]]$ code at $p = 10^{-3}$, at most nine at $p = 3\times10^{-3}$, and none for the $[[144,12,12]]$ code.
The hard decisions of non-converged probes thus stay in $L^{(0)}$ for a sizable part of the correctly decoded shots and rarely for failures.
We nevertheless keep the strict rule in the rest of this work, because the hard decision of a non-converged probe depends on the iteration at which the decoder stops (Section~\ref{sec:impl}), and we do not count a probe that has not converged as a return.

\section{Discussion}\label{sec:discussion}

In this work we introduce parity push post-selection (PPPS) for quantum low-density parity-check (qLDPC) codes, which penalizes the logical sector of the initial solution for each logical observable and accepts a shot only if every re-decoding still returns the initial logical class.
One slack variable and one auxiliary check added to the Tanner graph push the decoder toward the logical sector opposite to that of the initial solution.
Under DML decoding, all probes return exactly when the logical gap over the $2^K$ logical classes exceeds $\lambda$ [Eq.~\eqref{eq:threshold}], and $\lambda \to \infty$ recovers the forced run of FG.
Like FG, PPPS needs only $K$ re-decodings instead of one decoding in each of the $2^K$ logical classes.
Unlike FG, it keeps the initial solution feasible under the push, so a probe that fails to converge becomes a signal for rejection.

With Relay-BP, PPPS lowers the LER of the $[[72,12,6]]$ code at $p = 10^{-3}$ and of the $[[144,12,12]]$ code at $p = 3\times10^{-3}$ by about $50\times$ and $75\times$, respectively, at rejection rates near $1\%$.
At $p = 3\times10^{-3}$ the $[[72,12,6]]$ code needs a larger rejection rate than at $p=10^{-3}$, and PPPS lowers its LER by about $43\times$ at a rejection rate of $20\%$.
It reaches LERs below the error floor of FG in all three settings.
Against AR, PPPS stays within a factor of two in LER on the $[[72,12,6]]$ code at $p = 10^{-3}$ and needs about half the rejection rate for the same LER at $p = 3\times10^{-3}$.
On the $[[144,12,12]]$ code its LER is about five times lower than that of AR at rejection rates near $1\%$.
PPPS also runs on plain BP and BP+OSD without modification, and the rejection rate that it needs for a given LER depends on how often the probes of each decoder return on correctly decoded shots.
A relaxed rule that reads the hard decisions of non-converged probes lowers the rejection rate further at almost the same LER.

The rejection rates near $1\%$ at which PPPS reaches these reductions add little to the cost of a quantum protocol.
Every rejected shot must be prepared again, so a step with rejection rate $r$ takes $1/(1-r)$ attempts on average~\cite{lee2026efficient}, which is about $1.01$ at $r = 1\%$.
A protocol that restarts whenever one of its $N$ post-selected steps rejects succeeds with probability $(1-r)^N$, so the cost compounds with depth, and a low $r$ keeps this growth slow.
Post-selection at such rates can cost far less than enlarging the code~\cite{lee2026efficient}.
The classical cost equals that of FG, since both run $K$ re-decodings with the same budget (Section~\ref{sec:cost}).
Our probes use a smaller budget than the baseline (Table~\ref{tab:decoders}), so the probe stage adds only a fraction of the baseline latency (Section~\ref{sec:setup}).
A larger probe budget would raise this cost, and we expect it to lower the rejection rate because fewer correctly decoded shots would abstain.

One may ask why we soften the constraint instead of making the forced run converge, as Ref.~\cite{wills2026forced} proposes with OSD~\cite{panteleev2021degenerate} or decimation~\cite{yao2024belief}.
Such a remedy turns post-selection into three stages: the initial decoding, the forced run and a post-processing stage.
The third stage runs whenever the forced run fails to converge, which happens on almost every correctly decoded shot (Section~\ref{sec:practical}), so most shots pass through all three stages.
PPPS needs only two stages, the initial decoding and the probes.
The initial solution keeps each probe feasible under the push, so the probe converges on almost every correctly decoded shot and needs no third stage.
PPPS therefore reaches below the floor of FG with one fewer stage than these remedies.
Its probes, however, still involve the same high-degree auxiliary check as the forced runs of FG.

Finally, we identify limitations of our strategy and simulations, together with directions for future work.

\emph{Heuristic decoders.}
The push strength $\lambda$ is an exact gap threshold only under DML decoding.
With BP-family decoders the failure probability bound of Eq.~\eqref{eq:pfail} does not hold, and we report only measured LERs.
A bound that relates the outcomes of practical probes to the logical gap would provide further theoretical justification for PPPS with practical decoders.

\emph{Probe budget.}
The rejection rate of PPPS depends on the probe budget, since most rejected correctly decoded shots are ones on which a probe abstains.
A larger budget would lower this cost, but it may also let some failures return, so its net effect on the trade-off remains to be measured.
The $[[144,12,12]]$ code uses a probe budget four times as large as that of the $[[72,12,6]]$ code, so our data do not separate the effect of the code from that of the budget.
While the relaxed rule of Fig.~\ref{fig:relaxed} recovers part of the cost, the probe budget remains to be optimized.

\emph{Converging forced runs.}
We have not compared PPPS with forced runs that converge through OSD or decimation.
Such a comparison would show what the soft penalty of PPPS adds beyond simply the stage that it saves.

\emph{Hardware.}
We assume that a probe iteration takes as long as a baseline iteration (Section~\ref{sec:cost}), which may not hold since the auxiliary check of high degree can lengthen an iteration.
The checks of $H$ connect to at most $242$ fault mechanisms, whereas each of the $K$ parallel probes needs a decoder instance with an auxiliary check that connects to up to $14{,}264$ fault mechanisms for the $[[144,12,12]]$ code.
How this check affects the latency of an iteration in hardware remains to be tested.

\emph{Extensions.}
A sliding-window version, with a push per window, would extend PPPS to real-time decoding in the spirit of Ref.~\cite{lee2026efficient}.
Our simulations cover only memory experiments, and the surgery-based logical measurements of the $[[144,12,12]]$ code on which FG has been tested~\cite{wills2026forced,yoder2025tour} are a natural next test.
FG and PPPS can also be combined, since FG rejects failures with a small gap at almost no cost when its forced run converges and PPPS rejects failures on which the forced run does not converge.
A hybrid that uses FG when the forced run converges and PPPS when it does not would keep the low-cost regime of FG and the floor-free regime of PPPS.

\section{Methods}
\subsection{Pseudocode of PPPS}\label{sec:alg}

Algorithm~\ref{alg:ppps} gives the procedure of Fig.~\ref{fig:overview} for one shot.
The probes do not depend on each other, so we set no order for them.

\begin{algorithm}[t]
\caption{Parity push post-selection (one shot)}
\label{alg:ppps}
\begin{algorithmic}[1]
\Require DEM $(H, A, \{p_i\})$, syndrome $\sigma$, push strength $\lambda$
\Ensure \textsc{Accept} with correction $e^{(0)}$, or \textsc{Reject}   
\State $e^{(0)} \gets \Call{Decode}{H, \sigma, \{p_i\}}$
\If{not converged}
    \State \Return \textsc{Reject}                                    
\EndIf
\State $L^{(0)} \gets A e^{(0)}$
\For{$j = 1, \dots, K$ (independently)}
    \State $H' \gets \begin{pmatrix} H & 0 \\ A_j & 1 \end{pmatrix}$,\quad
           $\sigma' \gets (\sigma,\, 1 \oplus L^{(0)}_j)$,\quad
           $p_z \gets 1/(1+e^{\lambda})$
    \State $\hat{e}^{(j)} \gets \Call{Decode}{H', \sigma', \{p_i\} \cup \{p_z\}}$
           \Comment{cold start}
    \If{not converged \textbf{or} $A \hat{e}^{(j)} \neq L^{(0)}$}
           \Comment{abstain or flip}
        \State \Return \textsc{Reject}
    \EndIf
\EndFor
\State \Return \textsc{Accept} with $e^{(0)}$
\end{algorithmic}
\end{algorithm}

\subsection{Logical gap and the failure probability}\label{sec:gap}       
We use the notation of Sections~\ref{sec:dem} and~\ref{sec:ml} to derive Eqs.~\eqref{eq:pfail} and~\eqref{eq:threshold}.

\emph{Derivation of Eq.~\eqref{eq:pfail}.}
We fix the syndrome $\sigma$.
The actual fault lies in logical class $c$ with probability $Z_c/\sum_{c'} Z_{c'}$, where $Z_c$ is the class mass of Section~\ref{sec:dem}.
A DML decoder returns the class of largest mass, so it fails exactly when the actual fault lies in any other class:
\begin{equation}
  \mathcal{P}(\mathrm{fail}\mid\sigma)
  = \frac{Z^{(2)}+Z_{\mathrm{rest}}}{Z^{(1)}+Z^{(2)}+Z_{\mathrm{rest}}}.
  \label{eq:pfail-raw}
\end{equation}
Here $Z^{(1)}\ge Z^{(2)}$ are the two largest class masses and $Z_{\mathrm{rest}}$ is the total mass of the remaining $2^K-2$ classes.
Dividing the numerator and the denominator by $Z^{(2)}$ and writing $g=\ln(Z^{(1)}/Z^{(2)})$ and $\alpha=Z_{\mathrm{rest}}/Z^{(2)}$ gives Eq.~\eqref{eq:pfail}, so the failure probability depends on $\sigma$ only through $g$ and $\alpha$.
Equation~\eqref{eq:pfail} is a statement about each shot, and the LER is its average over the accepted syndromes.
For a given shot the failure probability decreases as $g$ grows.
Accepting only shots with $g>\lambda$ therefore bounds the failure probability of each accepted shot by $(1+\alpha)/(e^{\lambda}+1+\alpha)$, where $\alpha$ is that of the shot.
Every remaining class has mass at most $Z^{(2)}$, so $0 \le \alpha \le 2^K-2$, and the bound is strongest for small $\alpha$.
When the two largest classes carry almost all of the mass, $\alpha\to 0$ and Eq.~\eqref{eq:pfail} reduces to
\begin{equation}
  \mathcal{P}(\mathrm{fail}\mid g,\alpha)
  = \frac{1+\alpha}{e^{g}+1+\alpha}
  \;\approx\; \frac{1}{e^{g}+1}.
  \label{eq:pfail-twoclass}
\end{equation}
Equation~\eqref{eq:pfail-twoclass} is exact for a code with a single logical observable such as the surface code, which has only two logical classes, and it then coincides with the expression of Ref.~\cite{chen2025scalable} for the free-energy difference under maximum-likelihood decoding.

\emph{Derivation of Eq.~\eqref{eq:threshold}.}
We assume that a DML decoder also performs the initial decoding, so that $L^{(0)}$ is the class of largest mass $Z^{(1)}$.
For the probe on observable $j$, the slack variable $z$ has prior probability $p_z = P(z{=}1) = 1/(1+e^{\lambda})$ and $P(z{=}0) = e^{\lambda}/(1+e^{\lambda})$, and the auxiliary check enforces $A_j e \oplus z = 1 \oplus L^{(0)}_j$.
The value of $z$ is therefore fixed by $e$ as $z = 1 \oplus L^{(0)}_j \oplus A_j e$, so that $z=1$ when $A_j e = L^{(0)}_j$ and $z=0$ when $A_j e \neq L^{(0)}_j$.
Since the prior factorizes as $P'(e,z) = P(e)P(z)$, every $e$ with $He=\sigma$ has probability $P(e)P(z{=}1)$ if $A_j e = L^{(0)}_j$ and $P(e)P(z{=}0)$ otherwise, which is Eq.~\eqref{eq:reweight}.
Every configuration in class $c$ satisfies $A_j e = c_j$ and receives the same factor, so summing Eq.~\eqref{eq:reweight} over each class gives Eq.~\eqref{eq:mass}.

The probe on observable $j$ returns when $L^{(0)}$ remains the class of largest mass after the push.
A competing class with $c_j = L^{(0)}_j$ receives the same factor as $L^{(0)}$, so its comparison with $L^{(0)}$ is unchanged and $L^{(0)}$ still wins.
Only the classes with $c_j \neq L^{(0)}_j$ can overtake $L^{(0)}$, which gives Eq.~\eqref{eq:return}.
Suppose first that $e^{-\lambda} Z^{(1)} > Z^{(2)}$.
Since $Z^{(2)}$ is the largest mass of any class other than $L^{(0)}$, every probe $j$ satisfies
\begin{equation}
  \max_{c:\, c_j \neq L^{(0)}_j} Z_c \;\le\; Z^{(2)} \;<\; e^{-\lambda} Z^{(1)},
\end{equation}
and all probes return.
Conversely, the class of mass $Z^{(2)}$ differs from $L^{(0)}$ in at least one observable $j^\ast$ and therefore lies in the opposite logical sector of observable $j^\ast$.
If that probe returns, Eq.~\eqref{eq:return} gives $e^{-\lambda} Z^{(1)} > Z^{(2)}$.
Taking logarithms, $\ln Z^{(1)} - \lambda > \ln Z^{(2)}$, that is, $g > \lambda$, which proves Eq.~\eqref{eq:threshold}.
Under DML decoding a probe either returns or flips; abstention arises only when a practical decoder fails to converge (Section~\ref{sec:practical}).

\subsection{Circuits, noise model and decoders}\label{sec:setup-methods}

We use the Stim memory circuits distributed with the Relay-BP implementation of Ref.~\cite{muller2025relay}.
These circuits follow the depth-8 syndrome cycle of Ref.~\cite{bravyi2024high}, with seven layers of CNOTs and one layer of measurement and initialization, and run $d$ noisy cycles followed by a noiseless readout.
Each circuit is a Z-basis memory experiment whose $K=12$ observables are the logical Z operators.
The noise is uniform circuit-level noise of strength $p$, with two-qubit depolarizing noise after every CNOT, single-qubit depolarizing noise on idle data qubits, and errors in every preparation and measurement.
We extract the DEM with Stim~\cite{gidney2021stim}.

Table~\ref{tab:decoders} lists the decoder parameters.
Relay-BP returns the lowest-weight solution among its converged legs, and we modify its implementation only to exclude the slack variable from the memory strengths (Section~\ref{sec:impl}).
BP+OSD and plain BP use the \texttt{ldpc} package~\cite{roffe2020decoding}.
A shot counts as non-converged when the decoder returns no solution that satisfies the syndrome, which never happens for BP+OSD.

\begin{table}[t]
\centering
\caption{Decoder parameters for each setting.
Where two values are given, the first applies to the baseline and the second to the probes, which also serve as the forced runs of FG and the re-decodings of AR.
All other parameters are shared.
BP+OSD was not run on the $[[72,12,6]]$ code at $p=10^{-3}$, and plain BP uses the BP parameters of BP+OSD without OSD.}
\label{tab:decoders}
\begin{tabular}{@{}lccc@{}}
\toprule
 & $[[72,12,6]]$ & $[[72,12,6]]$ & $[[144,12,12]]$ \\
 & $p=10^{-3}$ & $p=3\times10^{-3}$ & $p=3\times10^{-3}$ \\
\midrule
\multicolumn{4}{@{}l}{\emph{Relay-BP}} \\
\texttt{gamma0}                & $0.1$          & $0.1$          & $0.1$ \\
\texttt{pre\_iter}             & $80$           & $80$           & $80$ \\
\texttt{num\_sets}             & $300$ / $25$   & $300$ / $25$   & $300$ / $100$ \\
\texttt{set\_max\_iter}        & $60$           & $60$           & $60$ \\
\texttt{gamma\_dist\_interval} & $(-0.19,\,0.26)$ & $(-0.19,\,0.26)$ & $(-0.24,\,0.66)$ \\
\texttt{stop\_nconv}           & $100$          & $100$          & $100$ \\
\midrule
\multicolumn{4}{@{}l}{\emph{BP+OSD}} \\
\texttt{bp\_method}            & ---            & min-sum        & min-sum \\
\texttt{max\_iter}             & ---            & $100$          & $100$ \\
\texttt{ms\_scaling\_factor}   & ---            & $0.625$        & $0.625$ \\
\texttt{osd\_method}           & ---            & OSD-CS         & OSD-CS \\
\texttt{osd\_order}            & ---            & $10$           & $10$ \\
\bottomrule
\end{tabular}
\end{table}

\subsection{PPPS implementation}\label{sec:impl}

We construct the deformed check matrix $H'_j$ of Eq.~\eqref{eq:Hprime} for each observable $j$.
The auxiliary check carries the syndrome bit $1\oplus L^{(0)}_j$ and the slack variable $z$ carries the prior $p_z = 1/(1+e^{\lambda})$.
In Relay-BP every probe reuses the seed of the baseline run, so all fault mechanisms receive the same sequence of memory strengths $\gamma$ as in the baseline.
The probes therefore differ from the baseline only by the gadget and the decoder budget, not by new random $\gamma$ values.
The slack variable carries no memory strength, which keeps the push at exactly $\lambda$ on every leg.

Each probe is an independent decoder call determined by $(\sigma, j, \lambda)$ and the seed of the shot.
We accept a shot only when all $K$ probes return.
To save computation we stop evaluating a shot once one of its probes flips, since the shot is then rejected regardless of the remaining probes.
A rejected shot counts as a flip if any of its probes flips and as an abstention otherwise.

Because the probes reuse the baseline seed, the probe at $\lambda = 0$ is the baseline decoding with the gadget attached but without a push.
At $\lambda = 0$ the slack variable has prior LLR zero, so the auxiliary check sends no message to the fault mechanisms.
The probe nevertheless differs from the baseline in two ways.
First, its budget is smaller, so it misses solutions that the baseline finds only in later legs.
Second, Relay-BP ends a leg as soon as its hard decision satisfies every check, and the probe has one more check to satisfy.
The hard decision on $z$ follows only the message from the auxiliary check and can disagree with the parity of the fault mechanisms in $\mathrm{supp}(A_j)$ while the messages are still changing.
The probe then runs longer on a leg that the baseline would have ended.
Relay-BP carries the state at the end of each leg into the next leg as memory, so all later legs of the probe differ from those of the baseline.
The $\lambda = 0$ point therefore measures the cost of the smaller budget and of the extra check together.

BP+OSD and plain BP use the same gadget and the same decision rule.
Unlike Relay-BP, these decoders run the probes with exactly the settings of their baseline, so their $\lambda = 0$ points isolate the effect of the added check.
For plain BP this effect is absent and the $\lambda = 0$ point coincides with the baseline.
For BP+OSD the added check changes the OSD step, and the $\lambda = 0$ point already rejects $9.1\%$ of the failures and $0.04\%$ of the correctly decoded shots of the $[[72,12,6]]$ code at $p = 3\times10^{-3}$.

The relaxed rule of Section~\ref{sec:relaxed} is evaluated on the same probe runs.
Instead of counting a non-converged probe as an abstention, it reads the class of the hard decision at which the probe stopped.

\subsection{Forced-gap and argument-reweighting baselines}\label{sec:baselines}

\emph{Forced gap.}
For each observable $j$ FG re-decodes under a hard constraint, appending the row $A_j$ to $H$ with the syndrome bit $1\oplus L^{(0)}_j$ and no slack variable~\cite{wills2026forced}.
Every converged leg of the baseline run and of the $K$ forced runs contributes its solution, and we keep the lowest-weight solution found in each logical class, following Eqs.~(2)--(4) of Ref.~\cite{wills2026forced}.
FG approximates the logical gap by the weight difference between the two lightest of these solutions.
FG accepts a shot when its gap exceeds a threshold $T$, and we sweep $T$ to obtain the curve.
When the runs find solutions in only one logical class, which requires that none of the forced runs converges, the gap is set to infinity by convention and the shot is accepted.
With our decoder budget the forced runs do not converge on $99.6\%$ of the correctly decoded shots of the $[[72,12,6]]$ code at $p=10^{-3}$, on $94.6\%$ at $p=3\times10^{-3}$ and on $98.8\%$ of those of the $[[144,12,12]]$ code at $p=3\times10^{-3}$.

\emph{Argument reweighting.}
AR~\cite{xie2026simple} re-decodes the syndrome after raising the prior probability of every mechanism in the baseline solution to the power $b=1+z$, which is the ratio-test variant of Ref.~\cite{xie2026simple}.
If the second solution lies in $L^{(0)}$, AR applies the same reweighting to the mechanisms of the second solution and decodes a third time.
A shot is accepted only when all three solutions lie in the same class, which is the three-round logical-error criterion.
We sweep $z\in\{0.001, 0.01, 0.1, 0.3, 0.5, 1, 2, 4\}$, and each value gives one point in Fig.~\ref{fig:relay}.
A re-decoding that fails to converge counts as a rejection, as an abstention does in PPPS.

The forced runs of FG and the re-decodings of AR use the same Relay-BP parameters and seeds as the PPPS probes (Table~\ref{tab:decoders}), so the three methods start from the same baseline solution and differ only in how they re-decode.
Because they run with our decoder settings and on our failures, their absolute values differ from the original reports, and the comparison is valid only within the same shots and the same decoder.

\subsection{Sampling and statistics}\label{sec:stats}

\emph{Failures.}
Logical failures are rare at our operating points, so we identify them with a two-pass procedure for the $[[72,12,6]]$ code at $p=10^{-3}$ and the $[[144,12,12]]$ code at $p=3\times10^{-3}$.
The first pass decodes every shot with the baseline Relay-BP stopped at its first converged leg.
The second pass runs the full baseline only on the shots that the first pass decodes incorrectly, and the failures of the second pass form the set of $W$ failures.
A shot that the first pass decodes correctly but the full baseline would not is missed by this procedure, so $W$ and the baseline LER are lower bounds.
For the $[[72,12,6]]$ code at $p=3\times10^{-3}$ we run the full baseline on all $N_s$ shots.
Comparing it with the first pass on the same shots shows that the first pass misses $74$ of the $324$ failures in this setting.
We run every post-selection method on all identified failures.

\emph{Correctly decoded shots.}
Running the probes on every correctly decoded shot is infeasible, so we run each method on a sample of $S$ correctly decoded shots.
For the $[[72,12,6]]$ code at $p=10^{-3}$ we draw $20{,}000$ shots uniformly at random, and for the same code at $p=3\times10^{-3}$ we take every $20$th of the $200{,}000$ shots, which gives $10{,}000$; in both cases we then exclude failures and non-converged shots, leaving the $S$ of Table~\ref{tab:samples}.
For the $[[144,12,12]]$ code we take $10{,}000$ shots at regular intervals from the first $16.15$ million shots; none of the $10{,}000$ is a failure or a non-converged shot.
FG and AR run on $9{,}890$ of these shots and PPPS on $5{,}292$ of them.
Both subsets are fixed by the shot index alone and are therefore not correlated with any decoding outcome; the remaining runs were not completed within our compute budget.
BP+OSD and plain BP on the $[[72,12,6]]$ code use the same $10{,}000$ shots as Relay-BP at $p=3\times10^{-3}$, and plain BP is read from the BP stage of BP+OSD, so its sample keeps only the shots on which BP converges.
For BP+OSD on the $[[144,12,12]]$ code we sample $2{,}063$ of the $1.06$ million shots.
Table~\ref{tab:samples} lists $N_s, E, W$ and $S$ for every decoder and setting.

\begin{table}[t]
\centering
\caption{Numbers of shots for every decoder and setting.
$N_s$ is the number of simulated shots, $E$ the number on which the baseline decoder does not converge, $W$ the number of logical failures among the converged shots and $S$ the number of sampled correctly decoded shots on which the post-selection methods run.
Failures are identified with the two-pass procedure (two-pass) or by running the full baseline on every shot (full).
Where two values of $S$ are given, the first applies to PPPS and the second to FG and AR.
Plain BP finds no failure on the $[[144,12,12]]$ code, so we do not post-select it (Section~\ref{sec:decoders}).}
\label{tab:samples}
\begin{tabular}{@{}llrrrrl@{}}
\toprule
Decoder & Code, $p$ & $N_s$ & $E$ & $W$ & $S$ & Failures \\
\midrule
Relay-BP & $[[72,12,6]]$, $10^{-3}$ & $40{,}000{,}000$ & $53$ & $521$ & $19{,}999$ & two-pass \\
Relay-BP & $[[72,12,6]]$, $3\times10^{-3}$ & $200{,}000$ & $936$ & $324$ & $9{,}936$ & full \\
Relay-BP & $[[144,12,12]]$, $3\times10^{-3}$ & $28{,}200{,}000$ & $3{,}085$ & $378$ & $5{,}292$ / $9{,}890$ & two-pass \\
BP+OSD & $[[72,12,6]]$, $3\times10^{-3}$ & $200{,}000$ & $0$ & $1{,}088$ & $9{,}942$ & full \\
BP+OSD & $[[144,12,12]]$, $3\times10^{-3}$ & $1{,}060{,}000$ & $0$ & $497$ & $2{,}063$ & full \\
Plain BP & $[[72,12,6]]$, $3\times10^{-3}$ & $200{,}000$ & $80{,}841$ & $140$ & $5{,}981$ & full \\
Plain BP & $[[144,12,12]]$, $3\times10^{-3}$ & 1{,}060{,}000 & 738{,}970 & $0$ & --- & full \\
\bottomrule
\end{tabular}
\end{table}

\emph{Estimators.}
Without post-selection only the $E$ non-converged shots are rejected, so the baseline point has
\begin{equation}
\mathrm{LER}_{\rm base}=\frac{W}{N_s-E},\qquad \mathrm{reject}_{\rm base}=\frac{E}{N_s}.
\label{eq:baseline}
\end{equation}
At a given $\lambda$ let $W_{\rm ret}$ be the number of failures whose probes all return and $S_{\rm ret}$ the number of sampled correctly decoded shots whose probes all return.
We write
\begin{equation}
  m=\frac{W_{\rm ret}}{W},\qquad \rho_c=\frac{S_{\rm ret}}{S}
\end{equation}
for the fractions of failures and of correctly decoded shots that are accepted.
Scaling $\rho_c$ from the sample to all $N_s-W-E$ correctly decoded shots gives the LER per shot and the rejection rate,
\begin{equation}
  \mathrm{LER}_{\rm post}=\frac{Wm}{(N_s-W-E)\rho_c+Wm},\qquad
  \mathrm{reject}_{\rm post}=\frac{(N_s-W-E)(1-\rho_c)+W(1-m)+E}{N_s}.
  \label{eq:twofactor-exact}
\end{equation}
We report the logical error rate per round $1-(1-\mathrm{LER}_{\rm post})^{1/d}$, where the number of rounds equals the distance $d$~\cite{bravyi2024high}. The rejection rate $\mathrm{reject}_{\rm post}$ is the fraction of shots that are rejected, and we do not normalize it per round.
The same expressions apply to FG and AR with their own acceptance rules.

\emph{Confidence intervals.}
Every identified failure is decoded, so $m$ carries no sampling error and the statistical uncertainty of the rejection rate comes from $\rho_c$ alone.
We compute a $95\%$ Wilson interval~\cite{wilson1927probable} for $\rho_c$ and insert its endpoints into Eq.~\eqref{eq:twofactor-exact}.
The uncertainty of the logical error rate comes from the small number $W_{\rm ret}$ of accepted failures.
We compute a $95\%$ Garwood interval~\cite{garwood1936fiducial} for this Poisson count and apply the per-round conversion to its endpoints, which is valid because the conversion is monotonic.
These intervals do not include the failures that the two-pass procedure misses, so the true $W$ and LERs may be higher than those we report.

\emph{Two-factor approximation.}
Failures are a small fraction of the converged shots in every setting (at most $W/N_s=0.54\%$ for BP+OSD on the $[[72,12,6]]$ code at $p=3\times10^{-3}$).
We therefore drop $Wm$ from the denominator of the LER in Eq.~\eqref{eq:twofactor-exact} and $W(1-m)/N_s$ from the rejection rate, and replace $N_s-W-E$ by $N_s-E$ in both.
With Eq.~\eqref{eq:baseline} this gives
\begin{equation}
\mathrm{LER}_{\rm post}\approx \mathrm{LER}_{\rm base}\,\frac{m}{\rho_c},\qquad
\mathrm{reject}_{\rm post}\approx \frac{E}{N_s}+\Bigl(1-\frac{E}{N_s}\Bigr)(1-\rho_c).
\label{eq:twofactor}
\end{equation}
We keep the term $E/N_s$ because it is not small for plain BP, whose baseline already rejects $40\%$ of the shots.
When the probes return on most correctly decoded shots ($\rho_c$ close to $1$), Eq.~\eqref{eq:twofactor} separates the two factors: the LER after post-selection follows $m$ and the rejection rate follows $\rho_c$ (Section~\ref{sec:decoders}).

\subsection{Computational cost}\label{sec:cost}

\begin{table}[t]
\centering
\caption{Cost of PPPS and AR relative to the baseline decoding, with every decoding at its iteration budget.
Both settings of the $[[72,12,6]]$ code share one DEM.
The latencies assume a fixed time per iteration.}
\label{tab:cost}
\begin{tabular}{@{}lcc@{}}
\toprule
 & $[[72,12,6]]$ & $[[144,12,12]]$ \\
\midrule
$\mathcal{E}$ & $91{,}116$ & $391{,}320$ \\
$\bar{\mathcal{E}}'/\mathcal{E}$ & $1.024$ & $1.024$ \\
$I_{\rm base}$ / $I_{\rm probe}$ & $18{,}080$ / $1{,}580$ & $18{,}080$ / $6{,}080$ \\
\midrule
\multicolumn{3}{@{}l}{\emph{PPPS}} \\
Operations, Eq.~\eqref{eq:cost-ops} & $2.07$ & $5.13$ \\
Latency, sequential, Eq.~\eqref{eq:cost-seq} & $2.05$ & $5.04$ \\
Latency, parallel, Eq.~\eqref{eq:cost-par} & $1.09$ & $1.34$ \\
\midrule
\multicolumn{3}{@{}l}{\emph{AR}} \\
Operations and latency & $1.17$ & $1.67$ \\
\bottomrule
\end{tabular}
\end{table}

We measure the number of operations ($C$) of one decoding by the number of BP iterations ($I$) times the number of edges ($\mathcal{E}$) of its Tanner graph, which equals the number of nonzero entries of $H$.
For iterative message-passing decoding, the number of iterations and the number of edges per iteration set the decoding complexity~\cite{sason2004complexity,sason2009bounds}.
Min-sum BP updates each check and each variable in a time linear in its degree~\cite{chen2005reduced}, so the remaining constant cancels in the ratio between PPPS and the baseline.
For the latency ($T$), we follow Ref.~\cite{muller2025relay} and count BP iterations.
An FPGA implementation completes each iteration in a fixed number of clock cycles~\cite{maurer2025realtimedecodinggrosscode,maurya2025fpgatailoredalgorithmsrealtimedecoding}, so $T$ is proportional to $I$.

The probe on observable $j$ decodes $H'_j$, which has $\mathcal{E}'_j=\mathcal{E}+|\mathrm{supp}(A_j)|+1$ edges.
All probes share one budget $I_{\rm probe}$, so the number of operations $C_{\rm PPPS}$ of PPPS relative to that of the baseline decoding $C_{\rm base}$ is given by Eq.~\eqref{eq:cost-ops}, where $\bar{\mathcal{E}}'$ is the mean of $\mathcal{E}'_j$ over the $K$ observables.
The latency of PPPS relative to that of the baseline follows Eq.~\eqref{eq:cost-seq} when the probes run in sequence and Eq.~\eqref{eq:cost-par} when they run in parallel.
\begin{align}
\frac{C_{\rm PPPS}}{C_{\rm base}} &= 1+K\,\frac{I_{\rm probe}}{I_{\rm base}}\,\frac{\bar{\mathcal{E}}'}{\mathcal{E}}, \label{eq:cost-ops}\\
\frac{T_{\rm seq}}{T_{\rm base}} &= 1+K\,\frac{I_{\rm probe}}{I_{\rm base}}, \label{eq:cost-seq}\\
\frac{T_{\rm par}}{T_{\rm base}} &= 1+\frac{I_{\rm probe}}{I_{\rm base}}. \label{eq:cost-par}
\end{align}

We take each decoding at its iteration budget.
A Relay-BP decoding runs at most $\texttt{num\_sets}+1$ legs: leg $0$ is the pre-iteration pass with \texttt{pre\_iter} iterations, and legs $1$ to \texttt{num\_sets} run \texttt{set\_max\_iter} iterations each.
This bound gives $18{,}080$ iterations for the baseline and $1{,}580$ and $6{,}080$ for the probes of the $[[72,12,6]]$ and $[[144,12,12]]$ codes.
Table~\ref{tab:cost} lists $\mathcal{E}$, $\bar{\mathcal{E}}'/\mathcal{E}$ and the resulting ratios.

FG adds the same row without the slack variable and runs its forced runs with the probe budget, so its cost equals that of PPPS.
AR leaves the Tanner graph unchanged and re-decodes at most twice~\cite{xie2026simple}, which gives $1.17$ and $1.67$ times the operations of the baseline.
Each round of AR reweights the error model by the output of the previous round, so its re-decodings cannot run in parallel and its latency grows by the same factor.
Two assumptions enter these numbers.
The baseline stops once $100$ legs converge, whereas the probes never reach this count and always use their full budget.
The bound on $I_{\rm base}$ therefore underestimates the ratio, which rises from $5.1$ to $7.0$ on the $[[144,12,12]]$ code with the recorded number of baseline legs.
The latency also assumes that a probe iteration takes as long as a baseline iteration, although the auxiliary check has a much higher degree than the checks of $H$ (Section~\ref{sec:discussion}).

\section*{Data availability}
The data that support the findings of this study are available from the corresponding author upon reasonable request.
\section*{Code availability}
The code used for the simulations in this study is available from the corresponding author upon reasonable request.
\bibliographystyle{quantum}
\bibliography{refs}

\begin{thebibliography}{10}

\bibitem{aliferis2008accuracy}
Panos Aliferis, Daniel Gottesman, and John Preskill.
\newblock ``Accuracy threshold for postselected quantum computation''.
\newblock Quantum Information and Computation {\bf 8}, 181--244~(2008).
\newblock  \href{http://arxiv.org/abs/quant-ph/0703264}{arXiv:quant-ph/0703264}.

\bibitem{knill2005quantum}
Emanuel Knill.
\newblock ``Quantum computing with realistically noisy devices''.
\newblock \href{https://dx.doi.org/10.1038/nature03350}{Nature {\bf 434}, 39--44}~(2005).

\bibitem{english2025thresholds}
Lucas~H. English, Dominic~J. Williamson, and Stephen~D. Bartlett.
\newblock ``Thresholds for postselected quantum error correction from statistical mechanics''.
\newblock \href{https://dx.doi.org/10.1103/nh49-52y2}{Physical Review Letters {\bf 135}, 120603}~(2025).

\bibitem{breuckmann2021quantum}
Nikolas~P. Breuckmann and Jens~Niklas Eberhardt.
\newblock ``Quantum low-density parity-check codes''.
\newblock \href{https://dx.doi.org/10.1103/PRXQuantum.2.040101}{PRX Quantum {\bf 2}, 040101}~(2021).

\bibitem{bravyi2024high}
Sergey Bravyi, Andrew~W. Cross, Jay~M. Gambetta, Dmitri Maslov, Patrick Rall, and Theodore~J. Yoder.
\newblock ``High-threshold and low-overhead fault-tolerant quantum memory''.
\newblock \href{https://dx.doi.org/10.1038/s41586-024-07107-7}{Nature {\bf 627}, 778--782}~(2024).

\bibitem{lee2026efficient}
Seok-Hyung Lee, Lucas~H. English, and Stephen~D. Bartlett.
\newblock ``Efficient post-selection for general quantum {LDPC} codes''.
\newblock \href{https://dx.doi.org/10.1038/s41534-026-01242-x}{npj Quantum Information {\bf 12}, 96}~(2026).

\bibitem{wills2026forced}
Adam Wills, Theodore~J. Yoder, and Isaac Chuang.
\newblock ``Forced gap post-selection for quantum {LDPC} codes and their operations''~(2026).
\newblock  \href{http://arxiv.org/abs/2605.20346}{arXiv:2605.20346}.

\bibitem{xie2026simple}
Haipeng Xie, Nobuyuki Yoshioka, Kento Tsubouchi, and Ying Li.
\newblock ``Simple, efficient, and generic post-selection decoding for {qLDPC} codes''.
\newblock \href{https://dx.doi.org/10.1103/sv11-f78l}{Physical Review Letters {\bf 137}, 110601}~(2026).

\bibitem{staples2026scalable}
J.~Wilson Staples, Winston Fu, and Jeff~D. Thompson.
\newblock ``Scalable postselection of quantum resources''~(2026).
\newblock  \href{http://arxiv.org/abs/2603.08697}{arXiv:2603.08697}.

\bibitem{bombin2024fault}
H{\'e}ctor Bomb{\'\i}n, Mihir Pant, Sam Roberts, and Karthik~I. Sahay.
\newblock ``Fault-tolerant postselection for low-overhead magic state preparation''.
\newblock \href{https://dx.doi.org/10.1103/PRXQuantum.5.010302}{PRX Quantum {\bf 5}, 010302}~(2024).

\bibitem{bravyi2005universal}
Sergey Bravyi and Alexei Kitaev.
\newblock ``Universal quantum computation with ideal {Clifford} gates and noisy ancillas''.
\newblock \href{https://dx.doi.org/10.1103/PhysRevA.71.022316}{Physical Review A {\bf 71}, 022316}~(2005).

\bibitem{akahoshi2026runtime}
Yutaro Akahoshi, Riki Toshio, Jun Fujisaki, Hirotaka Oshima, Shintaro Sato, and Keisuke Fujii.
\newblock ``Runtime reduction in lattice surgery utilizing timelike soft information''.
\newblock \href{https://dx.doi.org/10.1103/wz3p-sgdq}{Physical Review Research {\bf 8}, 033085}~(2026).

\bibitem{yoder2025tour}
Theodore~J. Yoder, Eddie Schoute, Patrick Rall, Emily Pritchett, Jay~M. Gambetta, Andrew~W. Cross, Malcolm Carroll, and Michael~E. Beverland.
\newblock ``Tour de gross: A modular quantum computer based on bivariate bicycle codes''~(2025).
\newblock  \href{http://arxiv.org/abs/2506.03094}{arXiv:2506.03094}.

\bibitem{hutter2014efficient}
Adrian Hutter, James~R. Wootton, and Daniel Loss.
\newblock ``Efficient {M}arkov chain {M}onte {C}arlo algorithm for the surface code''.
\newblock \href{https://dx.doi.org/10.1103/PhysRevA.89.022326}{Physical Review A {\bf 89}, 022326}~(2014).

\bibitem{gidney2025yoked}
Craig Gidney, Michael Newman, Peter Brooks, and Cody Jones.
\newblock ``Yoked surface codes''.
\newblock \href{https://dx.doi.org/10.1038/s41467-025-59714-1}{Nature Communications {\bf 16}, 4498}~(2025).
\newblock  \href{http://arxiv.org/abs/2312.04522}{arXiv:2312.04522}.

\bibitem{meister2024soft}
Nadine Meister, Christopher~A. Pattison, and John Preskill.
\newblock ``Efficient soft-output decoders for the surface code''~(2024).
\newblock  \href{http://arxiv.org/abs/2405.07433}{arXiv:2405.07433}.

\bibitem{smith2024mitigating}
Samuel~C. Smith, Benjamin~J. Brown, and Stephen~D. Bartlett.
\newblock ``Mitigating errors in logical qubits''.
\newblock \href{https://dx.doi.org/10.1038/s42005-024-01883-4}{Communications Physics {\bf 7}, 386}~(2024).

\bibitem{chen2025scalable}
Hongkun Chen, Daohong Xu, Grace~M. Sommers, David~A. Huse, Jeff~D. Thompson, and Sarang Gopalakrishnan.
\newblock ``Scalable accuracy gains from postselection in quantum error correcting codes''~(2025).
\newblock  \href{http://arxiv.org/abs/2510.05222}{arXiv:2510.05222}.

\bibitem{delfosse2021almost}
Nicolas Delfosse and Naomi~H. Nickerson.
\newblock ``Almost-linear time decoding algorithm for topological codes''.
\newblock \href{https://dx.doi.org/10.22331/q-2021-12-02-595}{Quantum {\bf 5}, 595}~(2021).

\bibitem{hillmann2025localized}
Timo Hillmann, Lucas Berent, Armanda~O. Quintavalle, Jens Eisert, Robert Wille, and Joschka Roffe.
\newblock ``Localized statistics decoding for quantum low-density parity-check codes''.
\newblock \href{https://dx.doi.org/10.1038/s41467-025-63214-7}{Nature Communications {\bf 16}, 8214}~(2025).

\bibitem{gidney2021stim}
Craig Gidney.
\newblock ``Stim: a fast stabilizer circuit simulator''.
\newblock \href{https://dx.doi.org/10.22331/q-2021-07-06-497}{Quantum {\bf 5}, 497}~(2021).

\bibitem{iyer2015hardness}
Pavithran Iyer and David Poulin.
\newblock ``Hardness of decoding quantum stabilizer codes''.
\newblock \href{https://dx.doi.org/10.1109/TIT.2015.2422294}{IEEE Transactions on Information Theory {\bf 61}, 5209--5223}~(2015).

\bibitem{kovalev2013quantum}
Alexey~A. Kovalev and Leonid~P. Pryadko.
\newblock ``Quantum {Kronecker} sum-product low-density parity-check codes with finite rate''.
\newblock \href{https://dx.doi.org/10.1103/PhysRevA.88.012311}{Physical Review A {\bf 88}, 012311}~(2013).

\bibitem{muller2025relay}
Tristan M{\"u}ller, Thomas Alexander, Michael~E. Beverland, Markus B{\"u}hler, Blake~R. Johnson, Thilo Maurer, and Drew Vandeth.
\newblock ``Improved belief propagation is sufficient for real-time decoding of quantum memory''~(2025).
\newblock  \href{http://arxiv.org/abs/2506.01779}{arXiv:2506.01779}.

\bibitem{panteleev2021degenerate}
Pavel Panteleev and Gleb Kalachev.
\newblock ``Degenerate quantum {LDPC} codes with good finite length performance''.
\newblock \href{https://dx.doi.org/10.22331/q-2021-11-22-585}{Quantum {\bf 5}, 585}~(2021).

\bibitem{roffe2020decoding}
Joschka Roffe, David~R. White, Simon Burton, and Earl~T. Campbell.
\newblock ``Decoding across the quantum low-density parity-check code landscape''.
\newblock \href{https://dx.doi.org/10.1103/PhysRevResearch.2.043423}{Physical Review Research {\bf 2}, 043423}~(2020).

\bibitem{fuentes2021degeneracy}
Patricio Fuentes, Josu Etxezarreta~Martinez, Pedro~M. Crespo, and Javier Garcia-Fr{\'i}as.
\newblock ``Degeneracy and its impact on the decoding of sparse quantum codes''.
\newblock \href{https://dx.doi.org/10.1109/ACCESS.2021.3089829}{IEEE Access {\bf 9}, 89093--89119}~(2021).

\bibitem{poulin2008iterative}
David Poulin and Yeojin Chung.
\newblock ``On the iterative decoding of sparse quantum codes''.
\newblock Quantum Information and Computation {\bf 8}, 987--1000~(2008).
\newblock  \href{http://arxiv.org/abs/0801.1241}{arXiv:0801.1241}.

\bibitem{raveendran2021trapping}
Nithin Raveendran and Bane Vasi{\'c}.
\newblock ``Trapping sets of quantum {LDPC} codes''.
\newblock \href{https://dx.doi.org/10.22331/q-2021-10-20-562}{Quantum {\bf 5}, 562}~(2021).

\bibitem{wilson1927probable}
Edwin~B. Wilson.
\newblock ``Probable inference, the law of succession, and statistical inference''.
\newblock \href{https://dx.doi.org/10.1080/01621459.1927.10502953}{Journal of the American Statistical Association {\bf 22}, 209--212}~(1927).

\bibitem{garwood1936fiducial}
Frank Garwood.
\newblock ``Fiducial limits for the {P}oisson distribution''.
\newblock \href{https://dx.doi.org/10.1093/biomet/28.3-4.437}{Biometrika {\bf 28}, 437--442}~(1936).

\bibitem{yao2024belief}
Hanwen Yao, Waleed Abu~Laban, Christian H{\"a}ger, Alexandre Graell~i Amat, and Henry~D. Pfister.
\newblock ``Belief propagation decoding of quantum {LDPC} codes with guided decimation''.
\newblock In Proc. IEEE Int. Symp. Inf. Theory (ISIT).
\newblock \href{https://dx.doi.org/10.1109/ISIT57864.2024.10619083}{Pages 2478--2483}.
\newblock ~(2024).
\newblock  \href{http://arxiv.org/abs/2312.10950}{arXiv:2312.10950}.

\bibitem{sason2004complexity}
Igal Sason and R{\"u}diger Urbanke.
\newblock ``Complexity versus performance of capacity-achieving irregular repeat-accumulate codes on the binary erasure channel''.
\newblock \href{https://dx.doi.org/10.1109/TIT.2004.828101}{IEEE Transactions on Information Theory {\bf 50}, 1247--1256}~(2004).

\bibitem{sason2009bounds}
Igal Sason and Gil Wiechman.
\newblock ``Bounds on the number of iterations for turbo-like ensembles over the binary erasure channel''.
\newblock \href{https://dx.doi.org/10.1109/TIT.2009.2018342}{IEEE Transactions on Information Theory {\bf 55}, 2602--2617}~(2009).

\bibitem{chen2005reduced}
Jinghu Chen, Ajay Dholakia, Evangelos Eleftheriou, Marc P.~C. Fossorier, and Xiao-Yu Hu.
\newblock ``Reduced-complexity decoding of {LDPC} codes''.
\newblock \href{https://dx.doi.org/10.1109/TCOMM.2005.852852}{IEEE Transactions on Communications {\bf 53}, 1288--1299}~(2005).

\bibitem{maurer2025realtimedecodinggrosscode}
Thilo Maurer, Markus B{\"u}hler, Michael Kr{\"o}ner, Frank Haverkamp, Tristan M{\"u}ller, Drew Vandeth, and Blake~R. Johnson.
\newblock ``Real-time decoding of the gross code memory with {FPGA}s''~(2025).
\newblock  \href{http://arxiv.org/abs/2510.21600}{arXiv:2510.21600}.

\bibitem{maurya2025fpgatailoredalgorithmsrealtimedecoding}
Satvik Maurya, Thilo Maurer, Markus B{\"u}hler, Drew Vandeth, and Michael~E. Beverland.
\newblock ``{FPGA}-tailored algorithms for real-time decoding of quantum {LDPC} codes''~(2025).
\newblock  \href{http://arxiv.org/abs/2511.21660}{arXiv:2511.21660}.

\end{thebibliography}

\section*{Acknowledgements}
The authors acknowledge helpful discussions with SangUk Lee and IlKwon Sohn. The authors also acknowledge helpful discussions with Adam Wills related to Ref.~\cite{wills2026forced}.

\section*{Funding}
H.J.K., G.S.M., and J.H. were supported by Institute for Information \& communications Technology Planning \& Evaluation (IITP) grant funded by the Korea government (MSIT) [No. 2020-0-00014, A Technology Development of Quantum OS for Fault-tolerant Logical Qubit Computing Environment].
S.H.L. was supported by the National Research Foundation of Korea (NRF) grant (RS-2026-25624201, RS-2026-25612364, RS-2026-25476454, RS-2024-00442710) funded by the Ministry of Science and ICT (MSIT) of the Korean government.

\section*{Author contributions}
H.J.K. conceived the idea and all authors contributed to developing it.
H.J.K. conducted the simulations and mainly wrote the manuscript.
G.S.M., S.H.L., and J.H. contributed helpful feedback and revisions.
All authors reviewed the manuscript.
\section*{Competing interests}
The authors declare no competing interests.

\ifSIappended
  \clearpage
  \input{si/si_setup}
  \begin{center}
{\Large Supplementary Information for}\\[4pt]
{\Large\bfseries Parity Push Post-Selection with Soft Parity Constraints for Quantum LDPC Codes}\\[8pt]
Hong Jun Kim, Gun Sik Min, Seok-Hyung Lee and Jun Heo
\end{center}

\snote{Origin of the rejections}{snote:origin}

This note collects the data behind the attributions in Sections~\ref{sec:relay} and~\ref{sec:relay144} of the main text and behind the interpretation in the Discussion.
Supplementary Tables~\ref{stab:ppps} and~\ref{stab:decoders} give the rejections of every setting by outcome.

\emph{Flips and abstentions.}
Failures are rejected mostly by flips, whereas correctly decoded shots are rejected mostly by abstentions.
At $\lambda=8$ a flip rejects $93\%$, $78\%$ and $94\%$ of the failures in the three panels of Fig.~\ref{fig:relay} of the main text, and an abstention rejects $4\%$, $21\%$ and $5\%$ of them; these shares change little with $\lambda$ (Supplementary Table~\ref{stab:ppps}, Supplementary Fig.~\ref{sfig:composition}).
At the same $\lambda$, abstentions on correctly decoded shots account for $0.28\%$ of the $0.32\%$ rejection rate of the $[[72,12,6]]$ code at $p=10^{-3}$, for $0.89\%$ of the $0.94\%$ of the $[[144,12,12]]$ code and for $17.5\%$ of the $20.2\%$ of the $[[72,12,6]]$ code at $p=3\times10^{-3}$.
Since failures are a small fraction of all shots, the rejection rate is set by the correctly decoded shots.
The cost of PPPS therefore comes almost entirely from abstentions on correctly decoded shots.

\emph{Convergence within the probe budget.}
Failures fail to converge within the probe budget far more often than correctly decoded shots, even without a push.
At $\lambda=0$ the probes abstain on $2.7\%$, $30\%$ and $15\%$ of the failures in the three settings, but only on $0.005\%$, $2.1\%$ and $0.02\%$ of the correctly decoded shots (Supplementary Table~\ref{stab:ppps}).
Supplementary Fig.~\ref{sfig:legs} shows the reason on the $[[144,12,12]]$ code.
The probes run legs $0$--$100$ and the baseline runs legs $0$--$300$ (Table~\ref{tab:decoders} and Section~\ref{sec:cost}).
On $56$ of the $378$ failures, the baseline decoder converges only after leg $100$.
The probes abstain on $57$ failures at $\lambda=0$, and these include all $56$.
Abstention without a push therefore marks the failures on which the decoder needs more legs than the probe budget allows.
Abstention is most frequent on the $[[72,12,6]]$ code at $p=3\times10^{-3}$, where the decoder needs more iterations to converge (Section~\ref{sec:relay} of the main text).
Since the probe budget is $25$ relay legs on the $[[72,12,6]]$ code and $100$ on the $[[144,12,12]]$ code (Table~\ref{tab:decoders} of the main text), the two codes cannot be compared directly.

\emph{Failures without a converged forced run.}
PPPS rejects most of the failures that FG accepts at its floor.
They lie among the failures on which none of the $K$ forced runs of FG converges.
FG accepts $18$ of these $21$ failures, $54$ of $76$ and $16$ of $22$ with an infinite gap~\cite{wills2026forced}, and finds a finite gap for the rest only from the solutions of the baseline run.
At $\lambda=8$ PPPS rejects $15$ of these $21$ failures, $71$ of $76$ and $21$ of $22$ in the three settings (Supplementary Table~\ref{stab:fgunconv}).
PPPS therefore falls below the FG floor.
Failures without a converged forced run are $23\%$ of all failures on the $[[72,12,6]]$ code at $p=3\times10^{-3}$, against $4\%$ at $p=10^{-3}$, so at the higher physical error rate PPPS falls below the floor with a smaller push.

\emph{Dependence on the decoder.}
At a given $\lambda$ the fraction of failures that return is small for every decoder, whereas the fraction of correctly decoded shots that return depends strongly on the decoder.
At $\lambda=4$ on the $[[72,12,6]]$ code at $p=3\times10^{-3}$, $7\%$, $4\%$ and $1\%$ of the failures return for Relay-BP, BP+OSD and plain BP, while $90\%$, $98\%$ and $34\%$ of the correctly decoded shots return (Supplementary Tables~\ref{stab:ppps} and~\ref{stab:decoders}).
BP+OSD never abstains, and plain BP rejects almost every correctly decoded shot by an abstention.
The first fraction sets the LER after post-selection and the second sets the rejection rate [Eq.~\eqref{eq:twofactor} of the main text], so the order of the three decoders in Fig.~\ref{fig:decoders} of the main text follows the fraction of correctly decoded shots that return.
BP+OSD does not abstain because OSD always returns a solution, so every rejection by BP+OSD is a flip.

\snote{Failures missed by the two-pass procedure}{snote:twopass}

The two-pass procedure identifies failures in two steps.
The first pass decodes every shot with the baseline decoder stopped at its first converged leg, and the second pass runs the full baseline only on the shots that the first pass decodes incorrectly (Section~\ref{sec:stats} of the main text).
A shot that the first pass decodes correctly but the full baseline decodes incorrectly is therefore never checked, and the procedure misses this failure.
For the $[[72,12,6]]$ code at $p=3\times10^{-3}$ we run the full baseline on every shot, which shows that the first pass misses $74$ of the $324$ failures.
Running PPPS on all of them shows that the missed failures are rejected more often than the other $250$.
Supplementary Table~\ref{stab:hidden} lists both groups at every $\lambda$.
At $\lambda=2$ PPPS rejects $96\%$ of the missed failures against $86\%$ of the others, and at $\lambda=8$ it rejects all of them against $98\%$ of the others.
FG also rejects all $74$ missed failures at its floor, against $196$ of the other $250$.

We use the two-pass procedure for the $[[72,12,6]]$ code at $p=10^{-3}$ and the $[[144,12,12]]$ code at $p=3\times10^{-3}$, so these settings likely contain failures that the procedure misses.
The true numbers of failures are then larger than those we identify, and we regard the baseline LERs and the LERs after post-selection of these settings as lower bounds.
Supplementary Table~\ref{stab:hidden} shows that PPPS rejects the missed failures of the $[[72,12,6]]$ code at $p=3\times10^{-3}$ more often than the other failures.
We therefore expect PPPS to reject most of the missed failures in the other two settings as well.
Running the full baseline on every shot of these settings, to see how PPPS rejects all of their failures, remains for future work.

\clearpage
\section*{Supplementary Figures}
\begin{figure}[htbp]\centering
\sifig{sfig1abc.pdf}
\caption{Composition of the rejections of PPPS with Relay-BP for (a) the $[[72,12,6]]$ code at $p=10^{-3}$, (b) the same code at $p=3\times10^{-3}$ and (c) the $[[144,12,12]]$ code at $p=3\times10^{-3}$, the three panels of Fig.~\ref{fig:relay} of the main text.
Each panel shows, against the push strength $\lambda$, the fraction of failures that PPPS rejects by a flip and by an abstention (solid lines), and the fraction of sampled correctly decoded shots that it rejects by a flip and by an abstention (dotted lines).
Failures are rejected mostly by flips and correctly decoded shots mostly by abstentions (Supplementary Note~\ref{snote:origin}).
The underlying counts are listed in Supplementary Table~\ref{stab:ppps}.}\label{sfig:composition}
\end{figure}
\begin{figure}[!tbp]\centering
\sifig[0.6\textwidth]{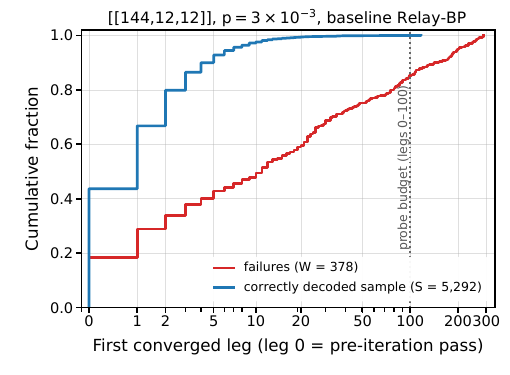}
\caption{Leg at which the baseline Relay-BP decoder first converges on the $[[144,12,12]]$ code at $p=3\times10^{-3}$, shown as the cumulative fraction of the $378$ failures and of the $5{,}292$ sampled correctly decoded shots.
Leg $0$ is the pre-iteration pass and legs $1$--$300$ are the relay legs (Section~\ref{sec:cost}).
The dotted vertical line marks leg $100$, the last leg of the probe budget; a shot on which the baseline converges only after this leg is likely to abstain in a probe.}\label{sfig:legs}
\end{figure}

\clearpage
\section*{Supplementary Tables}
\begin{table}[htbp]
\centering
\footnotesize
\caption{PPPS with Relay-BP (Fig.~\ref{fig:relay}).
For each push strength $\lambda$: failures whose probes all return ($W_{\rm ret}/W$) and failures rejected by a flip or an abstention; sampled correctly decoded shots whose probes all return ($S_{\rm ret}/S$) and those rejected by a flip or an abstention; the rejection rate per shot and the logical error rate per round with 95\% confidence intervals (Methods).}
\label{stab:ppps}
\resizebox{\ifdim\width>\linewidth\linewidth\else\width\fi}{!}{%
\begin{tabular}{@{}lllllll@{}}
\toprule
$\lambda$ & $W_{\rm ret}/W$ & \makecell[l]{Failures rejected\\ flip / abst} & $S_{\rm ret}/S$ & \makecell[l]{Correct rejected\\ flip / abst} & \makecell[l]{Rejection rate (\%)\\ {}[95\% CI]} & \makecell[l]{LER per round\\ {}[95\% CI]} \\
\midrule
\multicolumn{7}{@{}l}{\emph{$[[72,12,6]]$, $p=10^{-3}$}} \\
0 & $481/521$ & 26 / 14 & $19{,}998/19{,}999$ & 0 / 1 & 0.005 [0.001, 0.029] & 2.00 [1.8, 2.2] $\times10^{-6}$ \\
2 & $122/521$ & 378 / 21 & $19{,}992/19{,}999$ & 1 / 6 & 0.036 [0.018, 0.073] & 5.09 [4.2, 6.1] $\times10^{-7}$ \\
4 & $61/521$ & 445 / 15 & $19{,}988/19{,}999$ & 1 / 10 & 0.056 [0.032, 0.100] & 2.54 [1.9, 3.3] $\times10^{-7}$ \\
6 & $32/521$ & 473 / 16 & $19{,}973/19{,}999$ & 2 / 24 & 0.131 [0.090, 0.192] & 1.34 [0.91, 1.9] $\times10^{-7}$ \\
8 & $19/521$ & 483 / 19 & $19{,}936/19{,}999$ & 8 / 55 & 0.316 [0.248, 0.404] & 7.94 [4.8, 12] $\times10^{-8}$ \\
10 & $18/521$ & 485 / 18 & $19{,}861/19{,}999$ & 14 / 124 & 0.691 [0.586, 0.816] & 7.55 [4.5, 12] $\times10^{-8}$ \\
12 & $10/521$ & 494 / 17 & $19{,}761/19{,}999$ & 20 / 218 & 1.191 [1.050, 1.351] & 4.22 [2.0, 7.8] $\times10^{-8}$ \\
16 & $6/521$ & 497 / 18 & $19{,}175/19{,}999$ & 38 / 786 & 4.122 [3.855, 4.406] & 2.61 [0.96, 5.7] $\times10^{-8}$ \\
\midrule
\multicolumn{7}{@{}l}{\emph{$[[72,12,6]]$, $p=3\times10^{-3}$}} \\
0 & $175/324$ & 52 / 97 & $9{,}713/9{,}936$ & 15 / 208 & 2.773 [2.501, 3.081] & 1.50 [1.3, 1.7] $\times10^{-4}$ \\
1 & $66/324$ & 190 / 68 & $9{,}261/9{,}936$ & 55 / 620 & 7.348 [6.872, 7.856] & 5.94 [4.6, 7.6] $\times10^{-5}$ \\
2 & $38/324$ & 213 / 73 & $9{,}195/9{,}936$ & 66 / 675 & 8.022 [7.525, 8.552] & 3.44 [2.4, 4.7] $\times10^{-5}$ \\
3 & $33/324$ & 220 / 71 & $9{,}087/9{,}936$ & 82 / 767 & 9.104 [8.574, 9.667] & 3.03 [2.1, 4.2] $\times10^{-5}$ \\
4 & $22/324$ & 236 / 66 & $8{,}927/9{,}936$ & 105 / 904 & 10.710 [10.135, 11.316] & 2.05 [1.3, 3.1] $\times10^{-5}$ \\
6 & $10/324$ & 243 / 71 & $8{,}496/9{,}936$ & 156 / 1,284 & 15.026 [14.352, 15.728] & 9.81 [4.7, 18] $\times10^{-6}$ \\
8 & $6/324$ & 251 / 67 & $7{,}982/9{,}936$ & 213 / 1,741 & 20.169 [19.404, 20.957] & 6.26 [2.3, 14] $\times10^{-6}$ \\
12 & $3/324$ & 248 / 73 & $6{,}448/9{,}936$ & 340 / 3,148 & 35.512 [34.585, 36.450] & 3.88 [0.80, 11] $\times10^{-6}$ \\
16 & $2/324$ & 254 / 68 & $4{,}377/9{,}936$ & 433 / 5,126 & 56.225 [55.252, 57.192] & 3.81 [0.46, 14] $\times10^{-6}$ \\
\midrule
\multicolumn{7}{@{}l}{\emph{$[[144,12,12]]$, $p=3\times10^{-3}$}} \\
0 & $231/378$ & 90 / 57 & $5{,}290/5{,}292$ & 1 / 1 & 0.049 [0.022, 0.149] & 6.83 [6.0, 7.8] $\times10^{-7}$ \\
1 & $18/378$ & 348 / 12 & $5{,}285/5{,}292$ & 1 / 6 & 0.144 [0.076, 0.285] & 5.33 [3.2, 8.4] $\times10^{-8}$ \\
2 & $20/378$ & 345 / 13 & $5{,}279/5{,}292$ & 2 / 11 & 0.258 [0.156, 0.432] & 5.93 [3.6, 9.2] $\times10^{-8}$ \\
4 & $10/378$ & 351 / 17 & $5{,}281/5{,}292$ & 1 / 10 & 0.220 [0.128, 0.384] & 2.96 [1.4, 5.4] $\times10^{-8}$ \\
8 & $5/378$ & 354 / 19 & $5{,}243/5{,}292$ & 2 / 47 & 0.938 [0.713, 1.234] & 1.49 [0.48, 3.5] $\times10^{-8}$ \\
16 & $2/378$ & 355 / 21 & $4{,}814/5{,}292$ & 11 / 467 & 9.044 [8.301, 9.846] & 6.50 [0.79, 23] $\times10^{-9}$ \\
\bottomrule
\end{tabular}}
\end{table}

\begin{table}[htbp]
\centering
\footnotesize
\caption{FG with Relay-BP (Fig.~\ref{fig:relay}).
For each gap threshold $T$ (nats): sampled correctly decoded shots and failures whose gap is below $T$ and which FG therefore rejects, the rejection rate per shot and the logical error rate per round with 95\% confidence intervals.
A shot whose baseline run and forced runs find no solution of another class has an infinite gap and is always accepted; $T=\infty$ rejects every shot with a finite gap and gives the floor of the FG curve.}
\label{stab:fg}
\resizebox{\ifdim\width>\linewidth\linewidth\else\width\fi}{!}{%
\begin{tabular}{@{}lllll@{}}
\toprule
$T$ & \makecell[l]{Correct shots\\ rejected$/S$} & \makecell[l]{Failures\\ rejected$/W$} & \makecell[l]{Rejection rate (\%)\\ {}[95\% CI]} & \makecell[l]{LER per round\\ {}[95\% CI]} \\
\midrule
\multicolumn{5}{@{}l}{\emph{$[[72,12,6]]$, $p=10^{-3}$}} \\
0.5 & $0/19{,}999$ & $290/521$ & 0.001 [0.001, 0.020] & 9.63 [8.4, 11] $\times10^{-7}$ \\
1 & $0/19{,}999$ & $327/521$ & 0.001 [0.001, 0.020] & 8.08 [7.0, 9.3] $\times10^{-7}$ \\
2 & $0/19{,}999$ & $373/521$ & 0.001 [0.001, 0.020] & 6.17 [5.2, 7.2] $\times10^{-7}$ \\
3 & $0/19{,}999$ & $404/521$ & 0.001 [0.001, 0.020] & 4.88 [4.0, 5.8] $\times10^{-7}$ \\
4 & $0/19{,}999$ & $438/521$ & 0.001 [0.001, 0.020] & 3.46 [2.8, 4.3] $\times10^{-7}$ \\
6 & $1/19{,}999$ & $477/521$ & 0.006 [0.002, 0.030] & 1.83 [1.3, 2.5] $\times10^{-7}$ \\
8 & $9/19{,}999$ & $486/521$ & 0.046 [0.025, 0.087] & 1.46 [1.0, 2.0] $\times10^{-7}$ \\
10 & $17/19{,}999$ & $496/521$ & 0.086 [0.054, 0.137] & 1.04 [0.67, 1.5] $\times10^{-7}$ \\
12 & $23/19{,}999$ & $499/521$ & 0.116 [0.078, 0.174] & 9.18 [5.8, 14] $\times10^{-8}$ \\
16 & $39/19{,}999$ & $502/521$ & 0.196 [0.144, 0.268] & 7.93 [4.8, 12] $\times10^{-8}$ \\
24 & $75/19{,}999$ & $502/521$ & 0.376 [0.301, 0.471] & 7.95 [4.8, 12] $\times10^{-8}$ \\
$\infty$ & $92/19{,}999$ & $503/521$ & 0.461 [0.377, 0.565] & 7.53 [4.5, 12] $\times10^{-8}$ \\
\midrule
\multicolumn{5}{@{}l}{\emph{$[[72,12,6]]$, $p=3\times10^{-3}$}} \\
0.5 & $5/9{,}936$ & $70/324$ & 0.553 [0.524, 0.620] & 2.13 [1.9, 2.4] $\times10^{-4}$ \\
1 & $7/9{,}936$ & $112/324$ & 0.594 [0.558, 0.668] & 1.78 [1.5, 2.0] $\times10^{-4}$ \\
2 & $16/9{,}936$ & $155/324$ & 0.706 [0.644, 0.805] & 1.42 [1.2, 1.6] $\times10^{-4}$ \\
3 & $22/9{,}936$ & $185/324$ & 0.781 [0.706, 0.893] & 1.17 [0.98, 1.4] $\times10^{-4}$ \\
4 & $45/9{,}936$ & $203/324$ & 1.020 [0.906, 1.171] & 1.02 [0.85, 1.2] $\times10^{-4}$ \\
6 & $97/9{,}936$ & $232/324$ & 1.554 [1.380, 1.766] & 7.79 [6.3, 9.6] $\times10^{-5}$ \\
8 & $171/9{,}936$ & $244/324$ & 2.300 [2.064, 2.573] & 6.82 [5.4, 8.5] $\times10^{-5}$ \\
10 & $257/9{,}936$ & $251/324$ & 3.164 [2.871, 3.493] & 6.28 [4.9, 7.9] $\times10^{-5}$ \\
12 & $319/9{,}936$ & $258/324$ & 3.787 [3.460, 4.150] & 5.72 [4.4, 7.3] $\times10^{-5}$ \\
16 & $458/9{,}936$ & $261/324$ & 5.179 [4.786, 5.606] & 5.54 [4.3, 7.1] $\times10^{-5}$ \\
24 & $600/9{,}936$ & $266/324$ & 6.602 [6.153, 7.084] & 5.18 [3.9, 6.7] $\times10^{-5}$ \\
$\infty$ & $633/9{,}936$ & $270/324$ & 6.934 [6.473, 7.428] & 4.84 [3.6, 6.3] $\times10^{-5}$ \\
\midrule
\multicolumn{5}{@{}l}{\emph{$[[144,12,12]]$, $p=3\times10^{-3}$}} \\
0.5 & $0/9{,}890$ & $11/378$ & 0.011 [0.011, 0.050] & 1.08 [0.98, 1.2] $\times10^{-6}$ \\
1 & $0/9{,}890$ & $19/378$ & 0.011 [0.011, 0.050] & 1.06 [0.95, 1.2] $\times10^{-6}$ \\
2 & $0/9{,}890$ & $36/378$ & 0.011 [0.011, 0.050] & 1.01 [0.91, 1.1] $\times10^{-6}$ \\
3 & $0/9{,}890$ & $47/378$ & 0.011 [0.011, 0.050] & 9.78 [8.8, 11] $\times10^{-7}$ \\
4 & $0/9{,}890$ & $60/378$ & 0.011 [0.011, 0.050] & 9.40 [8.4, 10] $\times10^{-7}$ \\
6 & $1/9{,}890$ & $100/378$ & 0.021 [0.013, 0.069] & 8.22 [7.3, 9.2] $\times10^{-7}$ \\
8 & $1/9{,}890$ & $125/378$ & 0.021 [0.013, 0.069] & 7.48 [6.6, 8.5] $\times10^{-7}$ \\
10 & $2/9{,}890$ & $158/378$ & 0.032 [0.017, 0.085] & 6.50 [5.7, 7.4] $\times10^{-7}$ \\
12 & $2/9{,}890$ & $187/378$ & 0.032 [0.017, 0.085] & 5.65 [4.9, 6.5] $\times10^{-7}$ \\
16 & $4/9{,}890$ & $227/378$ & 0.052 [0.027, 0.116] & 4.46 [3.8, 5.2] $\times10^{-7}$ \\
24 & $10/9{,}890$ & $288/378$ & 0.113 [0.067, 0.198] & 2.66 [2.1, 3.3] $\times10^{-7}$ \\
$\infty$ & $136/9{,}890$ & $362/378$ & 1.387 [1.176, 1.636] & 4.79 [2.7, 7.8] $\times10^{-8}$ \\
\bottomrule
\end{tabular}}
\end{table}

\begin{table}[htbp]
\centering
\footnotesize
\caption{AR with Relay-BP (Fig.~\ref{fig:relay}).
For each reweighting strength $z$ ($b=1+z$): sampled correctly decoded shots and failures that AR rejects, the number of them rejected because a re-decoding did not converge (n.c.), the rejection rate per shot and the logical error rate per round with 95\% confidence intervals.}
\label{stab:ar}
\resizebox{\ifdim\width>\linewidth\linewidth\else\width\fi}{!}{%
\begin{tabular}{@{}lllllll@{}}
\toprule
$z$ & \makecell[l]{Correct shots\\ rejected$/S$} & \makecell[l]{of which\\ n.c.} & \makecell[l]{Failures\\ rejected$/W$} & \makecell[l]{of which\\ n.c.} & \makecell[l]{Rejection rate (\%)\\ {}[95\% CI]} & \makecell[l]{LER per round\\ {}[95\% CI]} \\
\midrule
\multicolumn{7}{@{}l}{\emph{$[[72,12,6]]$, $p=10^{-3}$}} \\
0.001 & $0/19{,}999$ & 0 & $281/521$ & 18 & 0.001 [0.001, 0.020] & 1.00 [0.88, 1.1] $\times10^{-6}$ \\
0.01 & $3/19{,}999$ & 3 & $315/521$ & 22 & 0.016 [0.006, 0.045] & 8.58 [7.5, 9.8] $\times10^{-7}$ \\
0.1 & $37/19{,}999$ & 35 & $470/521$ & 44 & 0.186 [0.136, 0.256] & 2.13 [1.6, 2.8] $\times10^{-7}$ \\
0.3 & $987/19{,}999$ & 947 & $518/521$ & 74 & 4.937 [4.645, 5.246] & 1.31 [0.27, 3.8] $\times10^{-8}$ \\
0.5 & $3{,}590/19{,}999$ & 3,529 & $519/521$ & 88 & 17.952 [17.426, 18.490] & 1.02 [0.12, 3.7] $\times10^{-8}$ \\
1.0 & $7{,}878/19{,}999$ & 7,804 & $520/521$ & 155 & 39.393 [38.718, 40.072] & 6.87 [0.17, 38] $\times10^{-9}$ \\
2.0 & $9{,}107/19{,}999$ & 9,037 & $520/521$ & 225 & 45.538 [44.849, 46.229] & 7.65 [0.19, 43] $\times10^{-9}$ \\
4.0 & $10{,}673/19{,}999$ & 10,605 & $520/521$ & 257 & 53.368 [52.676, 54.059] & 8.94 [0.23, 50] $\times10^{-9}$ \\
\midrule
\multicolumn{7}{@{}l}{\emph{$[[72,12,6]]$, $p=3\times10^{-3}$}} \\
0.001 & $348/9{,}936$ & 323 & $194/324$ & 109 & 4.045 [3.704, 4.423] & 1.13 [0.94, 1.3] $\times10^{-4}$ \\
0.01 & $406/9{,}936$ & 382 & $215/324$ & 111 & 4.636 [4.266, 5.041] & 9.53 [7.8, 11] $\times10^{-5}$ \\
0.1 & $2{,}082/9{,}936$ & 2,009 & $307/324$ & 187 & 21.444 [20.660, 22.250] & 1.80 [1.1, 2.9] $\times10^{-5}$ \\
0.3 & $6{,}795/9{,}936$ & 6,639 & $324/324$ & 234 & 68.587 [67.671, 69.488] & 0 [0, 9.8] $\times10^{-6}$ \\
0.5 & $8{,}804/9{,}936$ & 8,660 & $324/324$ & 263 & 88.679 [88.043, 89.285] & 0 [0, 2.7] $\times10^{-5}$ \\
1.0 & $9{,}601/9{,}936$ & 9,516 & $324/324$ & 293 & 96.650 [96.279, 96.985] & 0 [0, 9.2] $\times10^{-5}$ \\
2.0 & $9{,}706/9{,}936$ & 9,654 & $324/324$ & 311 & 97.700 [97.387, 97.976] & 0 [0, 1.3] $\times10^{-4}$ \\
4.0 & $9{,}763/9{,}936$ & 9,717 & $324/324$ & 311 & 98.270 [97.995, 98.507] & 0 [0, 1.8] $\times10^{-4}$ \\
\midrule
\multicolumn{7}{@{}l}{\emph{$[[144,12,12]]$, $p=3\times10^{-3}$}} \\
0.001 & $6/9{,}890$ & 6 & $292/378$ & 67 & 0.073 [0.040, 0.144] & 2.54 [2.0, 3.1] $\times10^{-7}$ \\
0.01 & $9/9{,}890$ & 9 & $306/378$ & 82 & 0.103 [0.060, 0.185] & 2.13 [1.7, 2.7] $\times10^{-7}$ \\
0.1 & $119/9{,}890$ & 113 & $354/378$ & 106 & 1.215 [1.019, 1.450] & 7.18 [4.6, 11] $\times10^{-8}$ \\
0.3 & $2{,}564/9{,}890$ & 2,547 & $378/378$ & 171 & 25.934 [25.080, 26.807] & 0 [0, 1.5] $\times10^{-8}$ \\
0.5 & $6{,}306/9{,}890$ & 6,282 & $378/378$ & 220 & 63.766 [62.813, 64.708] & 0 [0, 3] $\times10^{-8}$ \\
1.0 & $9{,}043/9{,}890$ & 9,032 & $378/378$ & 313 & 91.437 [90.869, 91.972] & 0 [0, 1.3] $\times10^{-7}$ \\
2.0 & $9{,}662/9{,}890$ & 9,656 & $378/378$ & 360 & 97.695 [97.380, 97.973] & 0 [0, 4.7] $\times10^{-7}$ \\
4.0 & $9{,}800/9{,}890$ & 9,794 & $378/378$ & 364 & 99.090 [98.883, 99.259] & 0 [0, 1.2] $\times10^{-6}$ \\
\bottomrule
\end{tabular}}
\end{table}

\begin{table}[htbp]
\centering
\footnotesize
\caption{PPPS with BP+OSD and plain BP at $p=3\times10^{-3}$ (Fig.~\ref{fig:decoders}).
Columns as in Supplementary Table~\ref{stab:ppps}; the Relay-BP values are those of Supplementary Table~\ref{stab:ppps}.}
\label{stab:decoders}
\resizebox{\ifdim\width>\linewidth\linewidth\else\width\fi}{!}{%
\begin{tabular}{@{}lllllll@{}}
\toprule
$\lambda$ & $W_{\rm ret}/W$ & \makecell[l]{Failures rejected\\ flip / abst} & $S_{\rm ret}/S$ & \makecell[l]{Correct rejected\\ flip / abst} & \makecell[l]{Rejection rate (\%)\\ {}[95\% CI]} & \makecell[l]{LER per round\\ {}[95\% CI]} \\
\midrule
\multicolumn{7}{@{}l}{\emph{BP+OSD, $[[72,12,6]]$}} \\
0 & $989/1088$ & 99 / 0 & $9{,}938/9{,}942$ & 4 / 0 & 0.090 [0.065, 0.152] & 8.27 [7.8, 8.8] $\times10^{-4}$ \\
0.25 & $241/1088$ & 847 / 0 & $9{,}872/9{,}942$ & 70 / 0 & 1.124 [0.978, 1.307] & 2.03 [1.8, 2.3] $\times10^{-4}$ \\
0.5 & $193/1088$ & 895 / 0 & $9{,}852/9{,}942$ & 90 / 0 & 1.348 [1.181, 1.553] & 1.63 [1.4, 1.9] $\times10^{-4}$ \\
1 & $152/1088$ & 936 / 0 & $9{,}829/9{,}942$ & 113 / 0 & 1.598 [1.409, 1.825] & 1.29 [1.1, 1.5] $\times10^{-4}$ \\
2 & $99/1088$ & 989 / 0 & $9{,}780/9{,}942$ & 162 / 0 & 2.115 [1.886, 2.382] & 8.43 [6.9, 10] $\times10^{-5}$ \\
4 & $43/1088$ & 1,045 / 0 & $9{,}706/9{,}942$ & 236 / 0 & 2.883 [2.604, 3.200] & 3.69 [2.7, 5.0] $\times10^{-5}$ \\
6 & $29/1088$ & 1,059 / 0 & $9{,}573/9{,}942$ & 369 / 0 & 4.221 [3.869, 4.609] & 2.52 [1.7, 3.6] $\times10^{-5}$ \\
8 & $15/1088$ & 1,073 / 0 & $9{,}386/9{,}942$ & 556 / 0 & 6.099 [5.666, 6.565] & 1.33 [0.75, 2.2] $\times10^{-5}$ \\
12 & $6/1088$ & 1,082 / 0 & $8{,}683/9{,}942$ & 1,259 / 0 & 13.136 [12.500, 13.800] & 5.76 [2.1, 13] $\times10^{-6}$ \\
\midrule
\multicolumn{7}{@{}l}{\emph{BP+OSD, $[[144,12,12]]$}} \\
0 & $478/497$ & 19 / 0 & $2{,}063/2{,}063$ & 0 / 0 & 0.002 [0.002, 0.188] & 3.76 [3.4, 4.1] $\times10^{-5}$ \\
0.25 & $16/497$ & 481 / 0 & $2{,}058/2{,}063$ & 5 / 0 & 0.288 [0.149, 0.611] & 1.26 [0.72, 2.0] $\times10^{-6}$ \\
0.5 & $6/497$ & 491 / 0 & $2{,}058/2{,}063$ & 5 / 0 & 0.289 [0.150, 0.612] & 4.73 [1.7, 10] $\times10^{-7}$ \\
1 & $5/497$ & 492 / 0 & $2{,}059/2{,}063$ & 4 / 0 & 0.240 [0.122, 0.544] & 3.94 [1.3, 9.2] $\times10^{-7}$ \\
2 & $5/497$ & 492 / 0 & $2{,}055/2{,}063$ & 8 / 0 & 0.434 [0.243, 0.809] & 3.95 [1.3, 9.2] $\times10^{-7}$ \\
4 & $1/497$ & 496 / 0 & $2{,}054/2{,}063$ & 9 / 0 & 0.483 [0.276, 0.873] & 7.90 [0.20, 44] $\times10^{-8}$ \\
6 & $2/497$ & 495 / 0 & $2{,}056/2{,}063$ & 7 / 0 & 0.386 [0.211, 0.745] & 1.58 [0.19, 5.7] $\times10^{-7}$ \\
8 & $1/497$ & 496 / 0 & $2{,}052/2{,}063$ & 11 / 0 & 0.580 [0.345, 0.999] & 7.91 [0.20, 44] $\times10^{-8}$ \\
12 & $0/497$ & 497 / 0 & $2{,}026/2{,}063$ & 37 / 0 & 1.840 [1.350, 2.508] & 0 [0, 3] $\times10^{-7}$ \\
\midrule
\multicolumn{7}{@{}l}{\emph{Plain BP, $[[72,12,6]]$}} \\
0 & $140/140$ & 0 / 0 & $5{,}981/5{,}981$ & 0 / 0 & 40.420 [40.420, 40.459] & 1.96 [1.6, 2.3] $\times10^{-4}$ \\
0.25 & $52/140$ & 61 / 27 & $5{,}278/5{,}981$ & 9 / 694 & 47.459 [46.988, 47.960] & 8.25 [6.2, 11] $\times10^{-5}$ \\
0.5 & $39/140$ & 78 / 23 & $5{,}068/5{,}981$ & 6 / 907 & 49.555 [49.026, 50.111] & 6.44 [4.6, 8.8] $\times10^{-5}$ \\
1 & $20/140$ & 90 / 30 & $4{,}852/5{,}981$ & 11 / 1,118 & 51.714 [51.136, 52.316] & 3.45 [2.1, 5.3] $\times10^{-5}$ \\
2 & $8/140$ & 99 / 33 & $3{,}704/5{,}981$ & 11 / 2,266 & 63.142 [62.415, 63.879] & 1.81 [0.78, 3.6] $\times10^{-5}$ \\
4 & $1/140$ & 107 / 32 & $2{,}062/5{,}981$ & 26 / 3,893 & 79.483 [78.761, 80.194] & 4.06 [0.10, 23] $\times10^{-6}$ \\
6 & $1/140$ & 107 / 32 & $926/5{,}981$ & 36 / 5,019 & 90.786 [90.227, 91.318] & 9.04 [0.23, 50] $\times10^{-6}$ \\
8 & $0/140$ & 104 / 36 & $124/5{,}981$ & 39 / 5,818 & 98.766 [98.532, 98.964] & 0 [0, 2.5] $\times10^{-4}$ \\
12 & $0/140$ & 103 / 37 & $2/5{,}981$ & 37 / 5,942 & 99.980 [99.927, 99.995] & 0 [0, 1.5] $\times10^{-2}$ \\
\bottomrule
\end{tabular}}
\end{table}

\begin{table}[htbp]
\centering
\footnotesize
\caption{PPPS with the relaxed rejection rule (Fig.~\ref{fig:relaxed}).
A probe that does not converge counts as a return when the class of its hard decision is $L^{(0)}$; otherwise the shot is rejected, and these rejections are listed as n.c.\ (not converged, hard decision outside $L^{(0)}$).
The other columns are as in Supplementary Table~\ref{stab:ppps}, which lists the strict-rule values.}
\label{stab:relaxed}
\resizebox{\ifdim\width>\linewidth\linewidth\else\width\fi}{!}{%
\begin{tabular}{@{}lllllll@{}}
\toprule
$\lambda$ & $W_{\rm ret}/W$ & \makecell[l]{Failures rejected\\ flip / n.c.} & $S_{\rm ret}/S$ & \makecell[l]{Correct rejected\\ flip / n.c.} & \makecell[l]{Rejection rate (\%)\\ {}[95\% CI]} & \makecell[l]{LER per round\\ {}[95\% CI]} \\
\midrule
\multicolumn{7}{@{}l}{\emph{$[[72,12,6]]$, $p=10^{-3}$}} \\
0 & $483/521$ & 26 / 12 & $19{,}998/19{,}999$ & 0 / 1 & 0.005 [0.001, 0.029] & 2.01 [1.8, 2.2] $\times10^{-6}$ \\
2 & $122/521$ & 378 / 21 & $19{,}998/19{,}999$ & 1 / 0 & 0.006 [0.002, 0.029] & 5.08 [4.2, 6.1] $\times10^{-7}$ \\
4 & $61/521$ & 445 / 15 & $19{,}992/19{,}999$ & 1 / 6 & 0.036 [0.018, 0.074] & 2.54 [1.9, 3.3] $\times10^{-7}$ \\
6 & $33/521$ & 473 / 15 & $19{,}984/19{,}999$ & 2 / 13 & 0.076 [0.047, 0.125] & 1.38 [0.95, 1.9] $\times10^{-7}$ \\
8 & $19/521$ & 483 / 19 & $19{,}958/19{,}999$ & 8 / 33 & 0.206 [0.153, 0.279] & 7.93 [4.8, 12] $\times10^{-8}$ \\
10 & $18/521$ & 485 / 18 & $19{,}911/19{,}999$ & 14 / 74 & 0.441 [0.359, 0.543] & 7.53 [4.5, 12] $\times10^{-8}$ \\
12 & $12/521$ & 494 / 15 & $19{,}855/19{,}999$ & 20 / 124 & 0.721 [0.613, 0.848] & 5.04 [2.6, 8.8] $\times10^{-8}$ \\
16 & $6/521$ & 497 / 18 & $19{,}514/19{,}999$ & 38 / 447 & 2.427 [2.222, 2.649] & 2.56 [0.94, 5.6] $\times10^{-8}$ \\
\midrule
\multicolumn{7}{@{}l}{\emph{$[[72,12,6]]$, $p=3\times10^{-3}$}} \\
0 & $184/324$ & 52 / 88 & $9{,}812/9{,}936$ & 15 / 109 & 1.778 [1.579, 2.015] & 1.56 [1.3, 1.8] $\times10^{-4}$ \\
1 & $67/324$ & 190 / 67 & $9{,}455/9{,}936$ & 55 / 426 & 5.407 [5.005, 5.844] & 5.90 [4.6, 7.5] $\times10^{-5}$ \\
2 & $41/324$ & 213 / 70 & $9{,}365/9{,}936$ & 66 / 505 & 6.320 [5.882, 6.792] & 3.65 [2.6, 4.9] $\times10^{-5}$ \\
3 & $36/324$ & 220 / 68 & $9{,}268/9{,}936$ & 82 / 586 & 7.293 [6.820, 7.799] & 3.24 [2.3, 4.5] $\times10^{-5}$ \\
4 & $23/324$ & 236 / 65 & $9{,}138/9{,}936$ & 105 / 693 & 8.599 [8.084, 9.147] & 2.10 [1.3, 3.1] $\times10^{-5}$ \\
6 & $13/324$ & 243 / 68 & $8{,}764/9{,}936$ & 156 / 1,016 & 12.345 [11.729, 12.990] & 1.24 [0.66, 2.1] $\times10^{-5}$ \\
8 & $6/324$ & 251 / 67 & $8{,}294/9{,}936$ & 213 / 1,429 & 17.049 [16.336, 17.787] & 6.03 [2.2, 13] $\times10^{-6}$ \\
12 & $3/324$ & 248 / 73 & $7{,}128/9{,}936$ & 340 / 2,468 & 28.711 [27.840, 29.599] & 3.51 [0.72, 10] $\times10^{-6}$ \\
16 & $2/324$ & 254 / 68 & $5{,}367/9{,}936$ & 433 / 4,136 & 46.324 [45.352, 47.299] & 3.11 [0.38, 11] $\times10^{-6}$ \\
\midrule
\multicolumn{7}{@{}l}{\emph{$[[144,12,12]]$, $p=3\times10^{-3}$}} \\
0 & $231/378$ & 90 / 57 & $5{,}290/5{,}292$ & 1 / 1 & 0.049 [0.022, 0.149] & 6.83 [6.0, 7.8] $\times10^{-7}$ \\
1 & $18/378$ & 348 / 12 & $5{,}286/5{,}292$ & 1 / 5 & 0.126 [0.064, 0.259] & 5.33 [3.2, 8.4] $\times10^{-8}$ \\
2 & $20/378$ & 345 / 13 & $5{,}281/5{,}292$ & 2 / 9 & 0.220 [0.128, 0.384] & 5.92 [3.6, 9.1] $\times10^{-8}$ \\
4 & $10/378$ & 351 / 17 & $5{,}284/5{,}292$ & 1 / 7 & 0.163 [0.089, 0.310] & 2.96 [1.4, 5.4] $\times10^{-8}$ \\
8 & $5/378$ & 354 / 19 & $5{,}255/5{,}292$ & 2 / 35 & 0.711 [0.520, 0.974] & 1.49 [0.48, 3.5] $\times10^{-8}$ \\
16 & $2/378$ & 355 / 21 & $4{,}910/5{,}292$ & 11 / 371 & 7.230 [6.563, 7.958] & 6.37 [0.77, 23] $\times10^{-9}$ \\
\bottomrule
\end{tabular}}
\end{table}

\begin{table}[htbp]
\centering
\footnotesize
\caption{PPPS on the failures on which none of the $K$ forced runs of FG converges (Supplementary Note~\ref{snote:origin}).
FG accepts $18$, $54$ and $16$ of them with an infinite gap.
For each $\lambda$ the number of these failures that PPPS rejects, with the number rejected by a flip and by an abstention in parentheses.
A dash marks a value of $\lambda$ that was not run for that setting.}
\label{stab:fgunconv}
\resizebox{\ifdim\width>\linewidth\linewidth\else\width\fi}{!}{%
\begin{tabular}{@{}llll@{}}
\toprule
$\lambda$ & \makecell[l]{$[[72,12,6]]$, $p=10^{-3}$\\ (of 21)} & \makecell[l]{$[[72,12,6]]$, $p=3\times10^{-3}$\\ (of 76)} & \makecell[l]{$[[144,12,12]]$, $p=3\times10^{-3}$\\ (of 22)} \\
\midrule
\multicolumn{4}{@{}l}{\emph{Relay-BP, failures rejected (flip / abst)}} \\
0 & 5 (0 / 5) & 43 (1 / 42) & 7 (1 / 6) \\
1 & -- & 62 (18 / 44) & 19 (14 / 5) \\
2 & 11 (1 / 10) & 65 (17 / 48) & 19 (12 / 7) \\
3 & -- & 67 (21 / 46) & -- \\
4 & 13 (4 / 9) & 69 (22 / 47) & 22 (13 / 9) \\
6 & 12 (2 / 10) & 69 (17 / 52) & -- \\
8 & 15 (4 / 11) & 71 (19 / 52) & 21 (14 / 7) \\
10 & 15 (2 / 13) & -- & -- \\
12 & 17 (4 / 13) & 73 (11 / 62) & -- \\
16 & 19 (5 / 14) & 74 (12 / 62) & 21 (11 / 10) \\
\bottomrule
\end{tabular}}
\end{table}

\begin{table}[htbp]
\centering
\footnotesize
\caption{Data for Supplementary Note~2.
Failures of the $[[72,12,6]]$ code at $p=3\times10^{-3}$ split into the $74$ that the first pass of the two-pass procedure misses and the other $250$, with the number that PPPS rejects at each $\lambda$ and at the FG floor.}
\label{stab:hidden}
\resizebox{\ifdim\width>\linewidth\linewidth\else\width\fi}{!}{%
\begin{tabular}{@{}lllll@{}}
\toprule
$\lambda$ & \makecell[l]{Missed failures rejected\\ (of 74)} & flip / abst & \makecell[l]{Other failures rejected\\ (of 250)} & flip / abst \\
\midrule
\multicolumn{5}{@{}l}{\emph{Relay-BP, $[[72,12,6]]$, $p=3\times10^{-3}$}} \\
0 & 49/74 (66.2\%) & 37 / 12 & 100/250 (40.0\%) & 15 / 85 \\
1 & 67/74 (90.5\%) & 65 / 2 & 191/250 (76.4\%) & 125 / 66 \\
2 & 71/74 (95.9\%) & 69 / 2 & 215/250 (86.0\%) & 144 / 71 \\
3 & 71/74 (95.9\%) & 67 / 4 & 220/250 (88.0\%) & 153 / 67 \\
4 & 73/74 (98.6\%) & 71 / 2 & 229/250 (91.6\%) & 165 / 64 \\
6 & 74/74 (100.0\%) & 72 / 2 & 240/250 (96.0\%) & 171 / 69 \\
8 & 74/74 (100.0\%) & 72 / 2 & 244/250 (97.6\%) & 179 / 65 \\
12 & 74/74 (100.0\%) & 69 / 5 & 247/250 (98.8\%) & 179 / 68 \\
16 & 74/74 (100.0\%) & 71 / 3 & 248/250 (99.2\%) & 183 / 65 \\
FG floor & 74/74 (100.0\%) & -- & 196/250 (78.4\%) & -- \\
\bottomrule
\end{tabular}}
\end{table}
\clearpage

\fi
\end{document}